\documentclass[preprint,aps,showpacs,nofootinbib]{revtex4}
\usepackage{epsfig}

\def\bea{\begin{eqnarray}}
\def\eea{\end{eqnarray}}
\def\bec{\begin{center}}
\def\ec{\end{center}}

\def\beq{\begin{equation}}
\def\eeq{\end{equation}}

\begin{document}

\begin{center}
{\Huge The Gaugino Code}
\vspace*{5mm} \vspace*{1cm}
\end{center}
\vspace*{5mm} \noindent
\centerline{\bf Kiwoon Choi ${}^{a}$ and  Hans  Peter Nilles
${}^{b}$} \vskip 1cm \centerline{\em ${{}^{a}}$ Department of
Physics, Korea Advanced Institute of Science and Technology}
\centerline{\em Daejeon 305-701, Korea} \vskip 5mm \centerline{\em
${{}^{b}}$ Physikalisches Institut, Universit\"at Bonn}
\centerline{\em Nussallee 12, D-53115 Bonn, Germany}

\vskip .9cm

\centerline{\bf Abstract}
\vskip .3cm
Gauginos might play a crucial role
in the search for supersymmetry at the Large Hadron Collider (LHC).
Mass predictions for gauginos are rather
robust and often related to the values of the gauge couplings.
We analyse the ratios of gaugino masses in
the LHC energy range for various schemes of supersymmetry
breakdown and mediation. Three distinct mass patterns emerge.

\vskip .3cm


\newpage

\section{introduction}

Soft breaking terms are the signals of the various schemes of
supersymmetry (SUSY) breakdown and its mediation to the
superpartners of the standard model (SM) particles \cite{reviews}. With the
upcoming experiments at LHC we might hope to identify these
superpartners and get information about their spectrum and
interactions. This in turn would allow us to infer the pattern of
the soft supersymmetry breaking terms. The crucial question then
concerns our ability to identify supersymmetry as the underlying
scheme and then draw conclusions about the
mechanism that is responsible for supersymmetry breakdown.

This could be a difficult task, as the relation between
superpartner spectra and the underlying scheme could be quite
complicated and model dependent, especially in the case of
incomplete experimental knowledge of the spectra. Strategies to
decode the spectra and determine the mechanism of
supersymmetry breakdown have to be
developed. Do there exist some model independent properties that
reveal special schemes? Can we make useful statements without the
knowledge of the heavy particle spectrum far beyond the TeV-scale?
In general, of course, we will not be able to answer these
questions in detail, but we might hope to identify some basic
characteristic patterns of soft terms.

From all the soft terms known, the gaugino masses have the simplest
form and appear to be the least model dependent, as e.g. compared to
the mass terms of squarks and sleptons. Therefore, an identification
of gaugino-like particles would be a first step in favour of a
potential supersymmetric interpretation of physics beyond the
standard model at the LHC.
In addition, gaugino masses $M_a$
$(a=1,2,3)$ are often simply related to the gauge coupling constants
$g_a$ of the SM gauge group $SU(3)\times SU(2)\times U(1)$. These
gauge coupling constants have been measured at the TeV-scale with
the (approximate) result: \bea g_1^2:g_2^2:g_3^2\,\simeq\, 1:2:6.
\eea In the minimal supersymmetric extension of the standard model
(MSSM) the renormalization group evolution of gaugino masses and
gauge coupling constants is related in a simple way: $M_a/g_a^2$
does not run at the one-loop level. In a basic scheme like
gravity mediation \cite{gravitymediation}
with a universal gaugino mass $M_{1/2}$ at the
grand unified (GUT) scale, i.e. mSUGRA scenario, the MSSM gaugino
masses at the TeV scale would thus obey the relation: \bea
\label{mgravity} \mbox{mSUGRA pattern:}\quad M_1:M_2:M_3\,\simeq\,
1:2:6 \eea
 which we call the mSUGRA pattern of low energy gaugino masses in
 the subsequent discussion.

Would such a pattern of gaugino masses uniquely point back to
gravity mediation with universal $M_{1/2}$ at the GUT-scale?
Or in other words,
how common is the mSUGRA pattern to various SUSY-breaking schemes?
Does it depend on the universality of gauge couplings and/or gaugino
masses at a large scale? What are other possible patterns of gaugino
masses which would result from a reasonable theoretical scheme
that could at the same time be clearly distinguished from the mSUGRA
pattern?

 In the present paper we would like to
address these questions and try to identify various patterns of
gaugino masses at the TeV-scale which might be obtained within a
reasonable theoretical framework. Our main goal is to see what kind
of information on SUSY breakdown can be extracted once one can
determine the low energy gaugino mass ratios by future collider
experiments. We shall find that only a few distinct patterns emerge,
some of them characteristic for a specific scheme, others shared by
quite different underlying schemes. It is encouraging to see that in
many cases the values of low energy gaugino masses are independent
of the particle spectrum at a large scale (like the GUT or
intermediate scale) and can therefore give us precious model
independent information.

Here is a summary of our results.
The  mSUGRA pattern (\ref{mgravity}) of low energy gaugino masses is
shared by many different schemes of SUSY breaking. These include, of
course, the mSUGRA-type SUSY-breaking scenarii realized in different
higher dimensional supergravity (SUGRA) or string theories with a
large string and compactification scales near $M_{GUT}\simeq 2\times
10^{16}$ GeV or $M_{Pl}\simeq 2.4\times 10^{18}$ GeV, e.g. the
dilaton/moduli-mediated SUSY breakdown in heterotic string/M theory
compactified on Calabi-Yau manifolds \cite{kaplunovskylouis,nilles},
flux-induced SUSY breakdown in Type IIB string theory \cite{ibanez},
as well as the gaugino mediation realized in higher dimensional
brane models \cite{gauginomediation}. These mSUGRA-type scenarios
either predict, or assume if necessary, the unification of gauge
couplings and/or gaugino masses at a large scale near $M_{GUT}$, and
they all give rise to the gaugino mass pattern (\ref{mgravity}) at
the TeV scale.

The scheme known as gauge mediation \cite{gaugemediation}, although
quite different from mSUGRA scenarii  in other aspects, also gives
rise to the low energy gaugino mass pattern (\ref{mgravity}) under
the assumption of gauge coupling unification at $M_{GUT}$. In fact,
even a broader class of different SUSY breaking schemes can lead to
the mSUGRA pattern. A nontrivial example of this kind is the large
volume compactification of Type IIB string theory discussed in
\cite{quevedo}. Although
the internal manifold of this compactification has an exponentially
large volume and the string scale  has an intermediate scale value
$M_{st}\sim 10^{11}$ GeV,  this scheme gives the mSUGRA pattern
(\ref{mgravity}) irrespective of whether or not the gauge couplings
are unified at the intermediate string scale.

There are two other simple patterns of low energy gaugino masses  distinct
from the mSUGRA pattern that emerge under suitable theoretical
assumptions that seem to be particularly appealing and well-motivated.
The first one is the one appearing in the scheme of anomaly
mediation \cite{anomalymediation}:
  \bea \label{anomalypattern}
\mbox{Anomaly pattern:}\quad  M_1:M_2:M_3\,\simeq\, 3.3:1:9, \eea
and the second one is mirage mediation \cite{mirage1}:
 \bea \label{mirage}
\mbox{Mirage pattern:}\quad M_1:M_2:M_3
\,\simeq\, (1+0.66\alpha):(2+0.2\alpha):(6-1.8\alpha),\eea where
$\alpha$ is a parameter of order unity that will be defined later.

The anomaly pattern in (\ref{anomalypattern}) requires that all
SUSY-breaking fields $X^I$ are sequestered from the visible sector.
Such sequestering might be naturally achieved in certain class of
theories with extra dimensions \cite{sequestering} or CFT sector
\cite{conformalsequestering}. Then the gaugino masses are dominated
by the SUGRA compensator-mediated contribution, such that \bea
\frac{M_a(\mu)}{g_a^2(\mu)}\,=\,\frac{b_a}{16\pi^2}m_{3/2},\eea
where $m_{3/2}$ is the gravitino mass and $b_a$ are the one-loop
beta-function coefficients at the scale $\mu$. Independently of the
UV structure at scales above TeV, $b_a=(33/5, 1, -3)$ at TeV if the
effective theory at TeV is given by the MSSM, which yields the
anomaly pattern  of low energy gaugino masses.

The mirage pattern in  (\ref{mirage}) is a kind of  hybrid between
the mSUGRA pattern and the anomaly pattern as it arises from
SUSY-breaking schemes in which the soft terms receive comparable
contributions from both moduli mediation and anomaly mediation. It
has been observed that such scheme is naturally realized in
KKLT-type moduli stabilization \cite{KKLT} and  its appropriate
generalizations \cite{LNR,PKKLT} which yield
\bea \label{miragepattern}
\frac{M_a(\mu)}{g_a^2(\mu)}\,=\,\left(1+\frac{\ln(M_{Pl}/m_{3/2})}{16\pi^2}
g_{GUT}^2b_a\alpha\right)\frac{M_{0}}{g_{GUT}^2},\eea
  where $M_0\sim 1$ TeV is a mass parameter characterizing the moduli mediation
  and  $\alpha={\cal O}(1)$ is a parameter
representing  the ratio of anomaly mediation to moduli mediation. In
this case, gaugino masses are {\it unified at a  mirage scale}
\cite{mirage2}: \bea
M_{\rm mir}=M_{GUT}\left(\frac{m_{3/2}}{M_{Pl}}\right)^{\alpha/2},
\eea and the resulting low energy values
take the mirage pattern in (\ref{mirage}) for  $g_{GUT}^2\simeq
1/2$.  In fact, the original KKLT-type moduli stabilization and some
of its generalizations  predict $\alpha\simeq 1$, although different
values of $\alpha={\cal O}(1)$ are possible in other
generalizations. As a consequence, the gaugino mass pattern
(\ref{mirage}) with $\alpha=1$ can be considered as a benchmark
point of the mirage pattern.

Another example of SUSY-breaking scheme leading to the mirage
pattern of gaugino masses is  deflected anomaly mediation \cite{DAM}
in which $M_a/g_a^2$ receive contributions from anomaly-mediation
and gauge-mediation of comparable size. In this case, $\alpha$
represents the ratio of anomaly
 to gauge mediation. Although it gives the same pattern of
low energy gaugino masses as mirage mediation, deflected anomaly
mediation  can be distinguished from mirage mediation  as it gives a
different pattern of soft scalar masses for a given value of
$\alpha$ \cite{mirage1,mirage2,DAM}.

Our scan of well motivated mediation scenarii thus leads to a remarkably
small number of distinct patterns of gaugino masses that could
be tested at the LHC.  At LHC, the cascade decays of gluino and
squarks are expected to provide information on the various
combinations of the gluino, squark, slepton, and neutralino masses
\cite{cascade}. Using the kinematic edges and thresholds of various
invariant mass distributions,  the gluino mass $M_3$ and the two
lightest neutralino masses $m_{\chi_1^0}$ and $m_{\chi_2^0}$ are
expected to be determined with a reasonable accuracy. In case that
$\chi_{1,2}^0$ are mostly the $SU(2)\times U(1)$ gauginos, this
would mean that the three MSSM gaugino masses are determined.  In the
other case that $\chi_1^0$ and $\chi_2^0$ correspond mostly to the
neutral Higgsinos, the dilepton invariant mass distribution of
$\chi_2^0\rightarrow \chi_1^0 ll^+$ shows a quite distinctive
feature \cite{kitano}, and then one might be able to extend the
kinematic analysis to determine the heavier neutralino masses
$\chi_{3,4}^0$ which are then expected to be the $SU(2)\times U(1)$
gaugino masses.
At any rate, under the assumption that the neutralino mixings are
not sizable, the mSUGRA pattern and the anomaly pattern predict
$M_{3}/m_{\chi_1^0}\simeq 6$ and $M_3/m_{\chi_1^0}\simeq 9$,
respectively. Thus, regardless of the nature of the LSP neutralino
$\chi_1^0$, they could be excluded if the data indicates the gluino
to LSP mass ratio significantly smaller than 6. On the other hand,
even when one finds the gluino to LSP mass ratio close to 6 or an
even larger value, one still needs to check the possibility of
Higgsino-like LSP in order to exclude the mirage pattern  predicting
$M_3/M_1$ significantly smaller than 6 for a positive $\alpha={\cal
O}(1)$.

With more data, any conclusion based on the gaugino masses
can then be checked, once candidates for squark and sleptons have
been identified. Mass predictions for squarks and sleptons, however,
show a stronger model dependence compared to the rather robust
gaugino mass patterns.
Of course, even in the case of gaugino masses we have to worry about
the possibility of a strong influence of physics at a high scale
(e.g. string threshold effects) that might be present in some of the
schemes \cite{binetruy,acharya} and obscure the 3 simple patterns
identified above. We shall describe these uncertainties once we
discuss the explicit scenarii in section III. Still we think that an
analysis of the gaugino masses is a most promising first step to
reveal the nature of the underlying scheme.

The organization of this paper is as follows. In the next section,
we discuss the general expression of $M_a/g_a^2$ in the context of
effective SUGRA. In section III, we consider various specific
examples including the scenarii which might be realized in the
context of  Type IIB string theory, heterotic string/M theory, and
also M theory compactified on a manifold with $G_2$ holonomy.
Section IV contains the summary and describes future strategies to
identify the underlying mechanism of supersymmetry breakdown.

\section{gaugino masses in 4d effective
supergravity}


To start with, let us consider 4D effective SUGRA defined at the
cutoff scale $\Lambda$ for the visible sector physics. This 4D
SUGRA might correspond to the low energy limit of compactified
string theory or brane model.
The Wilsonian effective action of the model at $\Lambda$ can be
written as \bea \label{effective} \int d^4\theta \,
CC^*\Big(-3e^{-K/3}\Big) + \left[ \int d^2\theta
\,\Big(\,\frac{1}{4}f_a W^{a\alpha}W^a_\alpha+ C^3 W\,\Big)+{\rm
h.c}\,\right],\eea where $C$ is the chiral compensator of 4D SUGRA,
$K$ is the K\"ahler potential, $W$ is the superpotential, and $f_a$
are holomorphic gauge kinetic functions. As usual, $K$ can be
expanded as \bea K&=&K_0(X_I,X_I^*)+Z_i(X_I,X_I^*)Q_i^*Q_i,\eea
where $Q_i$ are chiral matter superfields which have a mass lighter
than $\Lambda$ and are charged under the visible sector gauge group,
and $X_I$ are SUSY breaking (moduli or matter) fields which have
nonzero $F$-components $F^I$.

The running gauge couplings and gaugino masses at a scale $\mu$
below $\Lambda$ but above the next threshold scale $M_{\rm th}$ can
be determined by the 1PI gauge coupling superfield ${\cal F}_a(p^2)$
($M^2_{\rm th}<p^2<\Lambda^2$) which corresponds to the gauge
kinetic coefficient in the 1PI effective action on superpspace: \bea
\Gamma_{1PI}=\int d^4p\, d^4\theta \left(\, \frac{1}{4}{\cal
F}_a(p^2) W^a\frac{{\cal D}^2}{16 p^2}W^a+{\rm h.c}\,\right)\eea At
one-loop approximation, ${\cal F}_a$ is given by
\cite{kaplunovsky,bagger,arkani} \bea
\label{gaugecouplingsuperfield}{\cal F}_a(p^2)&=&{\rm
Re}(f^{(0)}_a)-\frac{1}{16\pi^2}(3C_a-\sum_iC_a^i)\ln\left(\frac{CC^*\Lambda^2}{p^2}\right)
\nonumber \\&&-\frac{1}{8\pi^2}\sum_i
C_a^i\ln\left(e^{-K_0/3}Z_i\right)+\frac{1}{8\pi^2}{\Omega}_a\eea
where $f_a^{(0)}$ are the tree-level gauge kinetic function, $C_a$
and $C_a^i$ are the quadratic Casimir of the gauge multiplets and
the matter representation $Q_i$, respectively. Here $\Omega_a$
contains the string and/or KK threshold corrections from heavy fields
at scales above $\Lambda$ as well as the (regularization
scheme-dependent) field-theoretic one-loop part:
$\frac{1}{8\pi^2}C_a\ln[{\rm Re}(f_a^{(0)})]$.
In the one-loop approximation, $\Omega_a$ are independent of the
external momentum $p^2$, thus independent of $C$ as a consequence of
the super-Weyl invariance. However $\Omega_a$ generically depend on
SUSY breaking  fields $X_I$, and a full determination of their
$X_I$-dependence requires a detailed knowledge of the UV physics
above $\Lambda$.

The running gauge couplings and gaugino masses at a
renormalization point $\mu$ ($M_{\rm th}<\mu<\Lambda$) are given
by
 \bea
\frac{1}{g_a^2(\mu)}&=&{\cal F}_a|_{C=e^{K_0/6},\,p^2=\mu^2},
\nonumber \\M_a(\mu)&=& F^A\partial_A\ln \left({\cal
F}_a\right)|_{{C=e^{K_0/6},\,p^2=\mu^2}},\eea where
$F^A=(F^C,F^I)$, $\partial_A=(\partial_C,\partial_I)$,
$C=e^{K_0/6}$ corresponds to the Einstein frame condition, and
\bea \frac{F^C}{C}=m^*_{3/2}+\frac{1}{3}F^I\partial_IK_0.\eea
 One then
finds \cite{kaplunovsky,bagger} \bea \label{highscaleratio}
\frac{1}{g_a^2(\mu)}&=& {\rm Re}(f^{(0)}_a)
-\frac{1}{16\pi^2}\left[(3C_a-\sum_i
C_a^i)\ln\left(\frac{\Lambda^2}{\mu^2}\right)\right.\nonumber \\
&&+\left.(C_a-\sum_iC_a^i)K_0+2\sum_iC_a^i\ln
Z_i\,\right]+\frac{1}{8\pi^2}\Omega_a,\nonumber \\
\frac{M_a(\mu)}{g_a^2(\mu)}&=&F^I\partial_I{\cal
F}_a+F^C\partial_C{\cal
F}_a \nonumber \\
&=& F^I\left[\frac{1}{2}\partial_I
f_a^{(0)}-\frac{1}{8\pi^2}\sum_i
C_a^iF^I\partial_I\ln(e^{-K_0/3}Z_i)+\frac{1}{8\pi^2}\partial_I\Omega_a\right]
\nonumber
\\
&&-\,\frac{1}{16\pi^2}(3C_a-\sum_iC_a^i)\frac{F^C}{C}. \eea
 Note that
$M_a/g_a^2$ do {\it not} run at one loop level, i.e. are independent of $\mu$, as
$M_a$ and $g_a^2$ have the same running behavior in the one-loop
approximation.

However, depending upon  the SUSY breaking scenario, the ratios
$M_a/g_a^2$ can receive important threshold corrections at lower
intermediate threshold scales $M_{\rm th}$. In fact, the expression
for $M_a/g_a^2$ in (\ref{highscaleratio}) is valid only for the
renormalization point between the high scale $\Lambda$ and and the
intermediate scale $M_{\rm th}$ where some of the particles
decouple. Let us now consider how  $M_a/g_a^2$ are modified by such
threshold effects at lower scale. To see this, we assume \bea
\{Q_i\}\equiv \{\Phi+\Phi^c,Q_x\},\eea and $\Phi+\Phi^c$ get a
supersymmetric mass of the order of $M_{\rm th}$,  while $Q_x$
remain to be massless at $M_{\rm th}$. Then $\Phi+\Phi^c$ can be
integrated out to derive the low energy parameters at scales below
$M_{\rm th}$. The relevant couplings of $\Phi+\Phi^c$ at $M_{\rm
th}$ can be written as \bea \int d^4\theta
CC^*e^{-K_0/3}\left(Z_\Phi\Phi^*\Phi+Z_{\Phi^c}\Phi^{c*}\Phi^c\right)
+\left(\int d^2\theta \,C^3\lambda_\Phi X_\Phi\Phi^c\Phi+{\rm
h.c}\right), \eea where $X_\Phi$ is assumed to have a vacuum
value\footnote{If $X_\Phi$ is not a superfield, but a parameter,
then $F^{X_\Phi}$ is obviously zero.} \bea \langle X_\Phi\rangle=
{M}_\Phi +\theta^2 F^{X_\Phi}.\eea Then the physical mass of
$\Phi+\Phi^c$ are given by \bea {\cal M}_\Phi=\lambda_\Phi\frac{C
X_\Phi}{\sqrt{e^{-2K_0/3}Z_\Phi Z_{\Phi^c}}},\eea yielding a
threshold correction
 to the gauge coupling superfield ${\cal F}_a$ as
 \bea
 \Delta {\cal F}_a(M_{\rm th})=
- \frac{1}{8\pi^2}\sum_{\Phi} C_a^\Phi \ln\left(\frac{{\cal
M}_\Phi
 {\cal M}_\Phi^*}{M_{\rm th}^2}\right).
 \eea
For $M_{\rm th}\sim {\cal M}_\Phi$, this gives rise to a threshold
correction  of ${\cal O}(1/8\pi^2)$ to $1/g_a^2$. In the leading log
approximation for  gauge couplings, such threshold corrections can
be ignored, therefore $1/g_a^2$ obeys the continuity condition at
$M_{\rm th}$: \bea \frac{1}{g_a^2(M_{\rm
th}^+)}=\frac{1}{g_a^2(M_{\rm th}^-)}, \eea where $M_{\rm th}^{\pm}$
denote the scale just above/below $M_{\rm th}$. On the other hand,
because $F^I$, $F^C$ and $F^{X^\Phi}$ can be quite different from
each other, the threshold correction to gaugino masses at $M_{\rm
th}$ can provide an important contribution to low energy gaugino
masses. For $\Delta {\cal F}_a$ given above, one easily finds that
the threshold correction to gaugino masses at $M_{\rm th}$ is given
by \bea &&M_a(M_{\rm th}^-)-M_a(M_{\rm th}^+)= g_a^2(M_{\rm
th})F^A\partial_A \Delta {\cal F}_a \nonumber
\\&=&-\frac{g_a^2(M_{\rm th})}{8\pi^2} \sum_\Phi C_a^\Phi
\left(\frac{F^C}{C}+\frac{F^{X_\Phi}}{M_\Phi}-
F^I\partial_I\ln(e^{-2K_0/3}Z_\Phi Z_{\Phi^c})\right) \eea Adding
this threshold correction to the result of (\ref{highscaleratio}),
we find
 \bea \left(\frac{M_a}{g_a^2}\right)_{M_{\rm th}^-}&=&
F^I\left[\frac{1}{2}\partial_I f_a^{(0)} -\frac{1}{8\pi^2}\sum_x
C_a^xF^I\partial_I\left(e^{-K_0/3}Z_x\right)+\frac{1}{8\pi^2}\partial_I\Omega_a\right]\nonumber
\\
&&-\frac{1}{8\pi^2}\sum_\Phi C_a^\Phi
\frac{F^{X_\Phi}}{M_\Phi}-\frac{1}{16\pi^2}(3C_a-\sum_xC_a^x)\frac{F^C}{C},
\eea where $\sum_x$ denotes the summation over $\{ Q_x\}$ which
remain as light matter fields at $M_{\rm th}^-$.

 One can repeat the above procedure, i.e. run down to the
lower threshold scale, integrate out the massive fields there, and
then include the threshold correction to gaugino masses until one
arrives at TeV scale. Then one finally finds
\bea\label{lowscaleratio} \left(\frac{M_a}{g_a^2}\right)_{\rm TeV}
=\tilde{M}_a^{(0)}+ \tilde{M}_a^{(1)}|_{\rm
anomaly}+\tilde{M}_a^{(1)}|_{\rm gauge}+\tilde{M}_a^{(1)}|_{\rm
string} \eea
 where \bea\label{components}
\tilde{M}_a^{(0)}&=&\frac{1}{2}F^I\partial_If_a^{(0)}, \nonumber \\
\tilde{M}_a^{(1)}|_{\rm anomaly} &=& \tilde{M}_a^{(1)}|_{\rm
conformal}+\tilde{M}_a^{(1)}|_{\rm Konishi}\nonumber \\
&=&\frac{1}{16\pi^2}b_a\frac{F^C}{C}-\frac{1}{8\pi^2}\sum_mC_a^mF^I\partial_I\ln(e^{-K_0/3}Z_m),
\nonumber \\
\tilde{M}_a^{(1)}|_{\rm gauge}&=& -\frac{1}{8\pi^2}\sum_\Phi
C_a^\Phi\frac{F^{X_\Phi}}{M_\Phi}, \nonumber \\
\tilde{M}_a^{(1)}|_{\rm string}&=&\frac{1}{8\pi^2}F^I\partial_I
\Omega_a.\eea
 Here $\sum_m$ denotes the
summation over the light matter multiplets $\{Q_m\}$ at the TeV
scale, $\sum_\Phi$ denotes the summation over the gauge messenger
fields $\Phi+\Phi^c$ which have a mass lighter than $\Lambda$ but
heavier than TeV, and \bea b_a=-3C_a+\sum_m C_a^m\eea are the
one-loop beta-function coefficients at TeV.  Obviously,
$\tilde{M}_a^{(0)}$ is the tree level value of $M_a/g_a^2$,
$\tilde{M}_a^{(1)}|_{\rm conformal}$ is the SUGRA
compensator-mediated one-loop contribution  determined by
the conformal anomaly of the effective theory at TeV scale
\cite{anomalymediation},  $\tilde{M}_a^{(1)}|_{\rm Konishi}$ is a
piece determined by the Konishi anomaly \cite{konishi},
$\tilde{M}_a^{(1)}|_{\rm gauge}$ are field theoretic gauge
thresholds  due to massive particles between $\Lambda$ and
the TeV scale, and
finally $\tilde{M}_a^{(1)}|_{\rm string}$ includes the
(UV-sensitive) string and/or KK thresholds at scales above $\Lambda$
as well as the (scheme-dependent) field theoretic one-loop piece
$\frac{1}{8\pi^2}C_aF^I\partial_I\ln[{\rm Re}(f_a^{(0)})]$.

Depending upon the SUSY breaking scenario, $M_a/g_a^2$ are dominated
by some of these five contributions. The stringy and KK thresholds
encoded in $\frac{1}{8\pi^2}\Omega_a$ are most difficult to compute
and highly model-dependent. In fact, this represents a potentially
uncontrollable contribution from high energy modes. If this part
gives an important contribution to $M_a/g_a^2$, no model independent
statements about the gaugino masses can be made. On the other hand,
the other parts can be reliably computed within the framework of 4D
effective theory under a reasonable assumption in many  SUSY
breaking scenarios. Note that the anomaly-related contribution
$\tilde{M}_a^{(1)}|_{\rm anomaly}$ is determined by the matter
contents at the TeV scale, while the other pieces require a
knowledge of physics at  scales higher than TeV.

We stress that (\ref{lowscaleratio}) is valid independently of the
matter content at scales above TeV, and thus is valid irrespective
of whether gauge couplings are unified or not at the initial cutoff
scale $\Lambda$. Also $\Lambda$ does not have to be close to
$M_{GUT}\simeq 2\times 10^{16}$ GeV. The result
(\ref{lowscaleratio}) can be applied  for the models with $\Lambda$
hierarchically lower than $M_{GUT}$.

Formulae (\ref{lowscaleratio}) and (\ref{components}) give the most
general description of gaugino masses and its origin from the
underlying schemes. This is our basic tool to analyse potential
candiates from future collider experiments. The SM gauge coupling
constants at TeV have been measured with the (approximate) result:
\bea g_1^2:g_2^2:g_3^2\simeq 1:2:6. \eea Once the gaugino mass
ratios at TeV are measured, the ratios of $M_a/g_a^2$ at TeV can be
experimentally determined, which will allow us to rule out many SUSY
breaking scenarios yielding $M_a/g_a^2$ different from the
experimental values. In the next section, we list the result of low
energy gaugino mass ratios for a variety of specific SUSY breaking
scenarii.

\section{specific examples}

\subsection{mSUGRA pattern}

\subsubsection{Gravity mediation}

The scheme of gravity mediation \cite{gravitymediation} with a
universal gaugino mass at $M_{GUT}$ is the one of the most popular
scenarii whose phenomenological consequences have been studied
extensively under the name of mSUGRA scenario. In this scheme,
$M_a/g_a^2$ are assumed to be universal at $M_{GUT}$, leading to the
mSUGRA pattern of gaugino masses at the TeV scale: \bea
M_1:M_2:M_3\,\simeq\, 1: 2: 6.\eea
 In the language of 4D effective SUGRA discussed in the previous section,
this amounts to assuming that the cutoff scale $\Lambda$ of 4D
effective SUGRA is close to $M_{Pl}$ or $M_{GUT}$, and $M_a/g_a^2$
of Eq.(\ref{lowscaleratio}) are dominated by the contribution
determined by the tree-level gauge kinetic function:
$\tilde{M}_a^{(0)}=\frac{1}{2}F^I\partial_If_a^{(0)}$, which is
assumed (or predicted) to be universal. Some interesting examples of
such scenario include dilaton/moduli mediation in heterotic
string/$M$-theory \cite{nilles}, flux-induced SUSY breakdown in Type
IIB string theory \cite{ibanez}, and gaugino mediation realized in
brane models \cite{gauginomediation}. In the following, we provide a
brief sketch of dilaton/moduli mediation in heterotic
string/$M$-theory and flux-induced SUSY breakdown in Type IIB string
theory.

\vskip 0.3cm {\it 1.1.  Dilaton/moduli-mediated SUSY breakdown in
heterotic string/$M$-theory:} \vskip 0.3cm

 The underlying UV theory of this
scheme is the 11D Horava-Witten theory \cite{HW}. At tree-level of
Horava-Witten theory  compactified on $CY\times S^1/Z_2$, the gauge
kinetic functions of the visible gauge fields take a universal form:
\bea f_a^{(0)}=S+\sum_i\beta_i T_i,\eea where ${\rm Re}(S)$ and
${\rm Re}(T_i)/[{\rm Re}(S)]^{1/3}$ ($i=1,..,h_{1,1}$) are
proportional to the CY volume and the length of the 11-th interval,
respectively, measured in 11D SUGRA unit, and $\beta_i$ are
topological numbers of order unity given by \cite{ck,witten}
\bea\beta_i=\frac{1}{8\pi^2}\int \omega_i\wedge(F\wedge
F-\frac{1}{2}R\wedge R),\eea where $\omega_i$ denote the basis of
harmonic two-forms on CY. In the region of moduli space in which
\bea {\rm Re}(S)={\cal O}(1),\quad {\rm Re}(T_i)={\cal O}(1)\eea for
the normalization of $S$ and $T$ determined by the above form of
$f_a^{(0)}$, the 11D SUGRA description provides a reliable
approximation for the UV theory and the corresponding
compactification scale  is close to $M_{GUT}\sim 2\times 10^{16}$
GeV \cite{witten}.  Under the assumption of dilaton/moduli
domination, $F^S/(S+S^*)$ and/or $F^i/(T_i+T_i^*)$ have a vacuum
value of ${\cal O}(m_{3/2})$. Then $M_a/g_a^2$ in
Eq.(\ref{lowscaleratio}) are dominated by the {\it universal}
tree-level contribution \cite{nilles}: \bea
\tilde{M}^{(0)}_a=\frac{1}{2}F^I\partial_If_a^{(0)}=\frac{1}{2}\left(
F^S+\sum_i\beta_i F^i\right)\eea since the other parts give a
subleading contribution of ${\cal O}(m_{3/2}/8\pi^2)$. Obviously,
then the resulting low energy gaugino masses take the mSUGRA
pattern.
Note that in this scenario the gauge coupling unification at
$M_{GUT}$ is predicted by the universal form of $f_a^{(0)}$, and the
universal gaugino masses at $M_{GUT}$ is also an automatic
consequence of the dilaton/moduli-dominated mediation scheme.

We note that as long as  $F^S/(S+S^*)={\cal O}(m_{3/2})$, the same
conclusion applies also  for the perturbative heterotic string limit
in which \bea
 {\rm Re}(S)\simeq 2,\quad {\rm
Re}(T_i)={\cal O}\left(\frac{1}{4\pi}\right).\eea In this region of
moduli space, the underlying UV theory corresponds to the weakly
coupled 10D heterotic string theory for which the tree level gauge
kinetic functions are given by $f_a^{(0)}=S$. For $M_a/g_a^2$ in
Eq.(\ref{lowscaleratio}),  the contribution $\frac{1}{2}\sum_i
\beta_i F^i$ which was identified as a part of the tree-level
contribution  $\tilde{M}_a^{(0)}=\frac{1}{2}F^I\partial_If_a^{(0)}$
in the heterotic $M$-theory limit should be considered as a part of the
string loop contribution $\tilde{M}_a^{(1)}|_{\rm
string}=\frac{1}{8\pi^2}F^I\partial_I\Omega_a$ in the perturbative
heterotic string limit. For the case that $F^S/(S+S^*)={\cal
O}(m_{3/2})$, $M_a/g_a^2$ in the perturbative heterotic string limit are
dominated again by the {\it universal} tree-level contribution
$\tilde{M}_a^{(0)}=\frac{1}{2}F^S$ since \bea
{\sum_i\beta_iF^i}\,\lesssim\, \frac{{\rm Re}(T_i)}{{\rm
Re}(S)}F^S\,\sim\, \frac{F^S}{8\pi},\eea and as a result the scheme
leads to the mSUGRA pattern of low energy gaugino masses
\cite{kaplunovskylouis}. On the other hand, in case that
$F^S/(S+S^*)=0$ while $F^i/(T_i+T^*_i)={\cal O}(m_{3/2})$, i.e.
moduli domination scenario, $M_a/g_a^2$ are determined mainly by
$\tilde{M}^{(1)}_a|_{\rm anomaly}$ and $\tilde{M}^{(1)}_a|_{\rm
string}$ which are generically non-universal, thus should be
discussed separately.

\vskip 0.3cm {\it 1.2. Flux-induced SUSY breakdown in Type IIB
string theory:} \vskip 0.3cm

In Type IIB string theory, the 3-form flux takes an imaginary
self-dual (ISD) value as a consequence of the equations of motion.
Such ISD flux contributes to the $F$-components of K\"ahler moduli
$T_i$, while giving a vanishing $F$-component for the dilaton and
complex structure moduli \cite{ibanez}. More explicitly, one finds
\bea
 \frac{F^i}{T_i+T^*_i}&=&-\frac{e^{K/2}}{T_i+T^*_i}\sum_j K^{i\bar{j}}(D_j W_{\rm flux})^*
 \nonumber \\
 &=&-e^{K/2}W^*_{\rm flux}\frac{\sum_j
 K^{i\bar{j}}\partial_{\bar{j}}K}{T_i+T_i^*}
 \eea
 for the flux-induced superpotential
 \bea
 W_{\rm flux}=\int (F_3-iSH_3)\wedge\Omega,
 \eea
where $F_3$ and $H_3$ are the RR and NS-NS fluxes, respectively, and
$\Omega$ is the holomorphic $(3,0)$-form of CY. At leading order in
the $\alpha^\prime$-expansion, the K\"ahler potential of K\"ahler
moduli obeys the no-scale relation
 \bea
 \label{noscale}
\sum_i(T_i+T_i^*)\partial_iK=-3,\quad \sum_j (T_j+T_j^*)K_{i\bar{j}}
=-\partial_iK\eea leading to {\it universal} K\"ahler moduli
$F$-components together with a vanishing $F$-component of the SUGRA
compensator: \bea \label{noscalepattern}\frac{F^i}{T_i+T_i^*}=
m^*_{3/2},\quad \frac{F^C}{C}=m^*_{3/2}+\frac{1}{3}F^i\partial_i
K=0.\eea

In fact, the above results of flux-induced SUSY breakdown have been
obtained in the limit in which $T_i$ are not stabilized yet. It has
been suggested that the overall volume modulus $T$ can be stabilized
by a competition between two small perturbative corrections to $K$,
while keeping (approximately) the no-scale structure
\cite{pstabilization}. Including the relevant higher order
corrections, the K\"ahler potential of $T$ takes the form:\bea
\label{corrected_Kahler}
K=-3\ln(T+T^*)+\frac{\xi_1}{(T+T^*)^{3/2}}-\frac{\xi_2}{(T+T^*)^2},\eea
where $\xi_1$ is the coefficient of higher order $\alpha^\prime$
correction and $\xi_2$ is the coefficient of string loop
corrections\footnote{In fact, string loop corrections to $K$
includes also a term of the form $X(S+S^*,{\cal Z},{\cal
Z}^*)/(T+T^*)$, where $X$ is a function of the string dilaton $S$
and complex structure moduli ${\cal Z}$. After $S$ and ${\cal Z}$
are fixed by flux, $X$ can be treated as a constant in the effective
theory of $T$. Then this correction of ${\cal O}(1/(T+T^*))$ can be
absorbed into a field redefinition $T+T^*\rightarrow
T+T^*-\frac{1}{3}X$, after which $K$ takes the form of
(\ref{corrected_Kahler}).}. If $\xi_1>0$ and $\xi_2>0$, one finds
that $T$ is stabilized with $F^T/(T+T^*)\simeq m_{3/2}^*$ and
$|F^C/C|\ll |m_{3/2}|$ \cite{pstabilization,choi}. However $\xi_1>0$
requires the Euler number $\chi=2(h_{1,1}-h_{2,1})$ of the
underlying CY orientifold to be positive as well. On the other hand,
most of interesting CY compactifications have nonzero $h_{2,1}$. In
particular, if one wishes to have a landscape of flux vacua which
might contain a state with nearly vanishing cosmological constant,
one typically needs $h_{2,1}={\cal O}(100)$ to accommodate a
sufficient number of independent 3-form fluxes. At the moment, it is
unclear how the above perturbative scheme of volume modulus
stabilization can be extended to the case with $h_{1,1}>1$,  while
keeping the flux-induced pattern  of SUSY breakdown maintained. It
remains to be seen whether such a scheme can be realized.

If such scheme of K\"ahler moduli stabilization exists, an
interesting feature of this flux-induced SUSY breakdown is that the
resulting $F^i/(T_i+T_i^*)$ have {\it universal} vacuum values at
leading order in the $\alpha^\prime$-expansion. To proceed, let us
suppose that the visible sector gauge fields live on $D7$ branes.
Then at leading order in the $\alpha^\prime$-expansion, the visible
sector gauge kinetic functions are generically given by \bea
f^{(0)}_a = \sum_i k_{ai}T_i,\eea where $k_{ai}$ are discrete
numbers of order unity.  Applying the universality of
$F^i/(T_i+T_i^*)$ and the vanishing $F^C$ to $M_a/g_a^2$ in
Eq.(\ref{lowscaleratio}) for
 this form of $f^{(0)}_a$,  one easily finds \bea
\left(\frac{M_a}{g_a^2}\right)_{\rm TeV}\,\simeq\,
\frac{m_{3/2}}{g_a^2(\Lambda)},\eea where $1/g_a^2(\Lambda)=
\sum_ik_{ai}{\rm Re}(T_i)$ for the cutoff scale $\Lambda$  of 4D
effective theory. Note that the gaugino masses at $\Lambda$ are
universal irrespective of the values of $k_{ai}$, i.e. irrespective
of whether or not the gauge couplings are unified at $\Lambda$. For
${\rm Re}(T_i)={\cal O}(1)$, $\Lambda$ is close to $M_{GUT}\sim
2\times 10^{16}$ GeV, and then it is reasonable to assume that
$g_a^2$ are unified at $\Lambda$: $g_a^2(\Lambda)\simeq g_{GUT}^2$.
Under this assumption of gauge coupling unification, $M_a/g_a^2$ at
the TeV scale are (approximately) universal, and thereby the low
energy gaugino masses take the mSUGRA pattern.

\subsubsection{Gauge mediation}

 In the gauge mediation scenario \cite{gaugemediation},  SUSY breakdown is
assumed to be mediated dominantly  by the loops of gauge-charged
messenger fields $\Phi+\Phi^c$ at a threshold scale $M_\Phi$ well
above TeV, so $M_a/g_a^2$ in Eq.(\ref{lowscaleratio}) are dominated
by the gauge-threshold contribution: \bea
\left(\frac{M_a}{g_a^2}\right)_{\rm
TeV}\,\simeq\,\tilde{M}_a^{(1)}|_{\rm gauge}=-
\frac{1}{8\pi^2}\sum_\Phi C_a^\Phi\frac{F^{X_\Phi}}{M_\Phi}.\eea
This gauge threshold correction to $M_a^2/g_a^2$ accompanies
additional running of gauge coupling constants over the scales
between $M_{GUT}$ and $M_\Phi$: \bea
 \Delta \left(\frac{1}{g_a^2}\right)=
- \frac{1}{8\pi^2}\sum_{\Phi} C_a^\Phi \ln\left(\frac{
M_{GUT}^2}{M_\Phi^2}\right).
 \eea
In order to maintain the successful gauge coupling unification of
the MSSM,  one usually assumes that the messenger fields
$\Phi+\Phi^c$ form a full $SU(5)$ multiplet, for which
$\tilde{M}_a^{(1)}|_{\rm gauge}$ are universal. The resulting low
energy gaugino masses then take the mSUGRA pattern, as a result
of the assumption of gauge coupling unification at $M_{GUT}$.

\subsubsection{Large volume compactification of Type IIB string
theory}

In models with an exponentially large compactification volume, the
messenger scale of SUSY breakdown is around the string scale
$M_{st}$ which is hierarchically lower than the 4D Planck scale. At
this moment, the only known example of  moduli stabilization giving
a large compactification volume  is the model of \cite{quevedo}
based on the following form of moduli K\"ahler potential and
superpotential: \bea
K&=&-2\ln\Big((T_b+T^*_b)^{3/2}-(T_s+T^*_s)^{3/2}+\xi\Big),
\nonumber
\\
W&=&w_0+Ae^{-aT_s},\eea where $T_b$ is the large 4-cycle K\"ahler
modulus for which the bulk CY volume is given by $V_{CY}\sim
(T_b+T_b^*)^{3/2}$ in the string unit with $M_{st}=1$, $T_s$ is the
K\"ahler modulus of small 4-cycle wrapped by D7 branes on which the
visible fields are assumed to live, and $w_0$ is assumed to be of
order unity.
 In the limit ${\rm Re}(T_b)\gg {\rm Re}(T_s)$, one
then finds \bea e^{a_{T_s}}\sim (T_b+T_b)^{3/2},\eea thus an
exponentially large volume when $aT_s\gg 1$, and also the following
pattern of $F$-components \cite{conlon}: \bea
\frac{F^{T_b}}{T_b+T_b}&=& m_{3/2}\left[1+{\cal O}\left(\frac{1}{(T_b+T_b^*)^{3/2}}\right)\right], \nonumber \\
\frac{F^{T_s}}{T_s+T_s^*}&\simeq&
\frac{m_{3/2}}{\ln(M_{Pl}/m_{3/2})},
\nonumber \\
\frac{F^C}{C}&=&m_{3/2}+\frac{1}{3}F^I\partial_IK_0\,=\, {\cal
O}\left(\frac{m_{3/2}}{(T_b+T_b^*)^{3/2}}\right),
\nonumber \\
F^S&=& F^U\,=\, 0,
 \eea
where $S$ and $U$ denote the string dilaton and complex structure
moduli, respectively. The string scale, 4D Planck scale and
$m_{3/2}$ are related as \cite{quevedo}\bea
\frac{M_{st}^2}{M_{Pl}^2}\,\sim\,\frac{m_{3/2}}{M_{Pl}}\,\sim\,
\frac{1}{(T_b+T_b^*)^{3/2}},\eea and $m_{3/2}\sim 10$ TeV (which
would give the sparticle mass $m_{\rm soft}\sim
m_{3/2}/\ln(M_{Pl}/m_{3/2})\sim 1$ TeV)  can be obtained for
$(T_b+T_b^*)^{3/2}\sim 10^{14}$ giving $M_{st}\sim 10^{11}$ GeV
\cite{quevedo}.

In the above $F^{T_b}$ originates mostly from  the 3-form flux
generating $w_0$ in the superpotential. On the other hand, $F^{T_s}$
receives contributions from both the flux and the non-perturbative
dynamics generating $Ae^{-aT_s}$, which dynamically cancel each
other, making $F^s$ suppressed by $1/\ln(M_{Pl}/m_{3/2})$. As the
K\"ahler potential of the large volume modulus $T_b$ takes the
no-scale form: \bea K_0=-3\ln(T_b+T_b^*)+{\cal
O}(1/(T_b+T_b)^{3/2}),\eea  $m_{3/2}$ in $F^C/C$ is cancelled
by as well $\frac{1}{3}F^{T_b}\partial_{T_b}K_0=-m_{3/2}$, making $F^C/C$
negligibly small.

Obviously, the visible sector gauge fields can not originate from
$D7$ branes wrapping the large 4-cycle since the corresponding gauge
couplings are too weak. For the gauge fields on $D7$ branes wrapping
the small 4-cycle, the tree-level gauge kinetic
functions\footnote{Such a scheme with $M_{st}\sim 10^{11}$ GeV
cannot accommodate (and is thus not constrained by) conventional
gauge coupling unification. The normalization of $U(1)_Y$ charge is
model-dependent and the $U(1)_Y$ gauge kinetic function can be
generalized to $f_{Y}=k_YT_s+h_YS$, where $k_Y$ is a generic
rational number. In such a case, the predicted Bino mass $M_1$
should be multiplied by $k_Y$.} are given by \cite{conlon} \bea
f^{(0)}_a=T_s+h_a S. \eea The corresponding gauge coupling
superfields ${\cal F}_a$ (see Eq.(\ref{gaugecouplingsuperfield}))
should be independent of the large 4-cycle volume ${\rm Re}(T_b)\sim
10^9$ \cite{conlon}, i.e. \bea
\partial_{T_b}(e^{K_0/3}Z_i)=\partial_{T_b}\Omega_a=0,\eea
so $F^{T_b}$ does not participate in the generation of the visible
sector  gaugino masses. Then, for the SUSY breaking pattern of large
volume compactification summarized above, $M_a/g_a^2$ are dominated
by the contribution from the {\it universal} tree-level gauge
kinetic function: \bea \left(\frac{M_a}{g_a^2}\right)_{\rm TeV}
=\frac{1}{2}F^I\partial_If_a^{(0)}=\frac{1}{2}F^{T_s},\eea leading
to the mSUGRA pattern of low energy gaugino masses:
 \bea
M_1:M_2:M_3\simeq 1:2:6. \eea

We stress that this result is obtained independently of the value of
the 4D cutoff scale $\Lambda\sim M_{st}$ at which ${\rm
Re}(f_a^{(0)})\simeq 1/g_a^2(\Lambda)$, and also of whether $g_a^2$
are unified  or not at $M_{st}$. For instance, for the case with
$M_{st}\sim 10^{11}$ GeV, $g_a^2(M_{st})$ might be unified or not,
depending upon whether or not there exist additional matter fields
other than those in the MSSM at scales between TeV and $M_{st}$. The
large volume compactification predicts that $M_a/g_a^2$ are
(approximately) universal regardless of the gauge coupling
unification near $M_{st}$ and also of the matter contents above TeV.
We also have \bea g_1^2:g_2^2:g_3^2\simeq 1:2:6,\eea  and thus the
mSUGRA pattern of gaugino masses  at TeV, regardless of whether
there exist extra matter states and also of whether $g_a^2$ are
unified at $M_{st}$.

\subsection{Mirage pattern}

In many compactifications of Type IIB theory with an explicit scheme
of K\"ahler moduli stabilization, SUSY is broken {\it not}
dominantly by flux, but by other effects such as the uplifting
mechanism. For instance, in KKLT stabilization of K\"ahler moduli
$T_i$ \cite{KKLT}, the flux-induced SUSY breaking $\propto D_iW_{\rm
flux}=(\partial_iK)W_{\rm flux}$ is {\it dynamically cancelled} by
the non-perturbative SUSY breaking $\propto D_iW_{\rm np}$
stabilizing $T_i$, thereby yielding a SUSY-preserving solution
$D_i(W_{\rm flux}+W_{\rm np})=0$ before the uplifting potential is
taken into account. Such a dynamical cancellation of flux-induced
SUSY breaking by moduli-stabilizing dynamics appears to be a
somewhat generic feature of moduli stabilization which admits a
supersymmetric configuration satisfying $D_iW=0$ since the admitted
supersymmetric configuration is always a stationary solution of the
SUGRA scalar potential\footnote{Of course, the perturbative
stabilization of $T$ by the K\"ahler potential
(\ref{corrected_Kahler}) is an exception since it does not admit
supersymmetric configuration in a region where the K\"ahler
potential (\ref{corrected_Kahler}) is reliable
\cite{pstabilization}. Note however that such perturbative
stabilization can be applied only for a quite limited situation as
we have noticed in the previous subsection.}.

Here we present three examples of SUSY breaking by uplifting
dynamics, the KKLT stabilization \cite{KKLT}, a partial-KKLT
stabilization \cite{PKKLT} and an uplift by F-terms of matter
superpotentials \cite{LNR}, all of which give rise to the mirage
pattern of low energy gaugino masses \cite{mirage1,mirage2} (but not
necessarily a mirage pattern for sfermion masses) if the visible
sector is assumed to live on $D7$ branes. We also present another
example of SUSY-breaking scenario, the deflected anomaly mediation
\cite{DAM}, which gives the same gaugino mass pattern although it is
quite different from the above three schemes in other aspects.

\subsubsection{KKLT with visible sector on D7 branes}

 KKLT stabilization of K\"ahler  moduli
$T_i$ in Type IIB string theory is based on the following features
of fluxed compactification: i) bulk geometry contains a warped
throat generated by flux, ii) SUSY is broken by a brane-localized
source stabilized at the IR end of warped throat, iii) there exist
non-perturbative dynamics, e.g. hidden gaugino condensation or
$D$-brane instantons, generating a non-perturbative superpotential of
the form $W_{\rm np}=\sum_i A_i e^{-a_i T_i}$. In this set-up,
SUSY-breaking at the IR end of the throat is sequestered from the
K\"ahler moduli and visible sector fields living in the bulk CY
located at the UV end of the throat, realizing the gravity dual of
the conformal sequestering scenario
\cite{sequestering,conformalsequestering}.  After integrating out
the heavy dilaton and complex structure moduli as well as the
massive degrees of freedom on SUSY-breaking brane, the effective
action of K\"ahler moduli is given by \bea {\cal L}_{\rm eff}=\int
d^4\theta \left[ -3CC^*\exp \left(-\frac{1}{3}K(T_i+T^*_i)\right)
-\theta^2\bar{\theta}^2 C^2C^{*2}{\cal P}_0\right] +\left(\int
d^2\theta \,C^3W +{\rm h.c}\right), \eea where $C$ is the SUGRA
compensator, the moduli K\"ahler potential $K$ obeys the no-scale
condition (\ref{noscale}) at leading order in the
$\alpha^\prime$-expansion, and \bea\label{ai}
W&=&w_0+\sum_i A_i e^{-a_i T_i}, \nonumber\\
 {\cal
P}_0&=&\mbox{constant},\eea where $w_0$ is a constant including the
flux-induced contribution. The uplifting operator
$\theta^2\bar{\theta}^2{\cal P}_0$ represents the low energy
consequence of the {\it sequestered} SUSY-breaking brane, and
thereby ${\cal P}_0$ is independent of $T_i$ and visible matter
fields.

In the Einstein frame with $C=e^{K/6}$, the above effective
lagrangian  gives the K\"ahler moduli potential: \bea V_{\rm
TOT}=V_F+V_{\rm lift}= e^K\left(K^{i\bar{j}}D_i W (D_j
W)^*-3|W|^2\right)+e^{2K/3}{\cal P}_0, \eea where $V_F$ is the
conventional $N=1$ SUGRA  potential and $V_{\rm lift}=e^{2K/3}{\cal
P}_0$ is the uplifting potential induced by  SUSY-breaking brane.
The K\"ahler moduli vacuum values and their $F$-components can be
computed by minimizing $V_{\rm TOT}$ under the condition $\langle
V_{\rm TOT}\rangle=0$. This can be done in two steps: first minimize
$V_F$ yielding a supersymmetric AdS solution of $D_i W=0$, and next
compute the small vacuum shifts $\delta T_i$ caused by $V_{\rm
lift}$ in a perturbative expansion in powers of
$1/\ln(M_{Pl}/m_{3/2})$ \cite{mirage1,mirage2,GKKLT}. The first step
yields the following moduli vacuum values:\bea \label{susysol} a_i
T_i= \ln(M_{Pl}/m_{3/2})\left[1+{\cal
O}\left(\frac{1}{\ln(M_{Pl}/m_{3/2})}\right)\right], \eea where the
$A_i$ of (\ref{ai}) have been assumed to be of order unity. As the
K\"ahler moduli masses at this supersymmetric solution are
significantly heavier than $m_{3/2}$, \bea m_{T_i}={\cal
O}\Big(m_{3/2}\ln(M_{Pl}/m_{3/2})\Big),\eea while $V_{\rm lift}$ has
a vacuum value of ${\cal O}(m_{3/2}^2)$ (in the unit with
$M_{Pl}=1$), the vacuum shift induced by $V_{\rm lift}$ is as small
as
 \bea
 \delta T_i\equiv \delta\phi_i+i\delta a_i
 ={\cal O}\left(\frac{1}{[\ln(M_{Pl}/m_{3/2})]^2}\right).\eea This makes the
computation of the moduli $F$-components in KKLT stabilization
rather straightforward.

To proceed, let us choose the field basis for which $w_0$ and $A_i$
take a real value. Since $V_{\rm lift}$ is a real function of real
variables $T_i+T_i^*$, it is obvious that $V_{\rm lift}$ does not
induce a tadpole of the axion component: $\delta a_i=0$. To compute
$\delta\phi_i$, let us expand $V_{\rm TOT}$ around the
supersymmetric solution $\vec{T}_0$ of $D_iW=0$: \bea
\label{expansion_V_TOT} V_{\rm TOT}&=&-3m_{3/2}^2(\vec{T}_0) +V_{\rm
lift}(\vec{T}_0)+2\partial_iV_{\rm lift}(\vec{T}_0)\delta\phi_i
+\left(m_\phi^2\right)_{ij}\delta\phi_i\delta\phi_j +..., \eea where
$\partial_iV_{\rm lift}=\partial V_{\rm lift}/\partial T_i$, and
\bea
\left(m_\phi^2\right)_{ij}&=&m_{3/2}^2a_ia_jK^{i\bar{j}}\partial_iK\partial_{\bar{j}}K
\left[1+{\cal
O}\left(\frac{1}{\ln(M_{Pl}/m_{3/2})}\right)\right].\eea
Upon ignoring the small corrections further suppressed by
$1/\ln(M_{Pl}/m_{3/2})$, minimizing $V_{\rm TOT}$ under the
condition $\langle V_{\rm TOT}\rangle=0$ leads to \bea
\sum_j\left(m_\phi^2\right)_{ij}\delta\phi_j=-\partial_iV_{\rm
lift}=-\frac{2}{3}V_{\rm lift}\partial_iK
=-2m_{3/2}^2\partial_iK,\eea and thus \bea \sum_j
a_ia_jK^{i\bar{j}}\partial_{\bar{j}}K\delta\phi_j=-2,\eea where we
have used that $V_{\rm lift}=e^{2K/3}{\cal P}_0$ for a constant
${\cal P}_0$, and also $\langle V_{\rm lift}\rangle\simeq
3m_{3/2}^2$ as required for $\langle V_{\rm TOT}\rangle\simeq 0$.
Applying the above result and (\ref{susysol}) to \bea
F^i&=&-e^{K/2}K^{i\bar{j}}(D_j W)^*\nonumber \\
&=& -m_{3/2} \sum_j a_jK^{i\bar{j}}\partial_{\bar{j}}K\delta
\phi_j\left[1+{\cal
O}\left(\frac{1}{\ln(M_{Pl}/m_{3/2})}\right)\right], \eea one finds
that $F^i/(T_i+T_i^*)$ are {\it universal} \cite{GKKLT}
 at leading order in
the expansion in $1/\ln(M_{Pl}/m_{3/2})$: \bea \frac{F^i}{T_i+T_i^*}
=\frac{m_{3/2}}{\ln(M_{Pl}/m_{3/2})}\equiv M_0.\eea  Note that the
universality of $F^i/(T_i+T^*_i)$ is obtained
independently of the detailed form of the  moduli K\"ahler
potential. The $F$-component of the SUGRA compensator $C$ in KKLT
stabilization is given by\bea
\frac{F^C}{C}=m_{3/2}+\frac{1}{3}\sum_iF^i\partial_iK\simeq
m_{3/2}.\eea

Applying the above pattern of $F$-components to $M_a/g_a^2$ in
Eq.(\ref{lowscaleratio}) for the case that visible gauge fields live
on $D7$ branes, i.e. for the case with \bea f^{(0)}_a=\sum_i
k_{ai}T_i,\eea we find that $M_a/g_a^2$ are dominated by the
tree-level contribution
$\tilde{M}_a^{(0)}=\frac{1}{2}F^i\partial_if_a^{(0)}$ and the
one-loop conformal anomaly contribution $\tilde{M}_a^{(1)}|_{\rm
conformal}$ which are comparable to each other:\bea
\left(\frac{M_a}{g_a^2}\right)_{\rm TeV}&=& \frac{1}{2}\sum_i
k_{ai}F^i+\frac{b_a}{16\pi^2}m_{3/2}\nonumber \\
& =&\left(1+ \frac{\ln(M_{Pl}/m_{3/2})}{16\pi^2}g_{GUT}^2
b_a\right)\frac{M_0}{g_{GUT}^2},\eea where we have assumed that
$g_a^2$ are unified at $M_{GUT}$: $\sum_ik_{ai}{\rm
Re}(T_i)={1}/{g_a^2(M_{GUT})}={1}/{g_{GUT}^2}$, and again  the
subleading parts suppressed by $1/\ln(M_{Pl}/m_{3/2})$ are ignored.
This corresponds to the mirage pattern (\ref{miragepattern}) with
$\alpha=1$, for which the low energy gaugino mass ratios are given
by \bea M_1:M_2:M_3\simeq 1:1.3:2.5. \eea

We note that a more general form of mirage pattern can be obtained
for instance when the gauge kinetic functions are generalized to
include higher order correction in the $\alpha^\prime$-expansion
\cite{CJ}, e.g. \bea f_a^{(0)}=\sum_i k_{i}T_i+hS,\eea where we have
assumed a {\it universal} form of $f_a$ for the gauge coupling
unification at $M_{GUT}$. The higher order term $hS$ in $f_a^{(0)}$
can be induced by a magnetic flux on $D7$ branes. Even when the
compactification radius $R$ is significantly bigger than the string
length scale $l_s=\sqrt{\alpha^\prime}$, so ${\rm Re}(S)/{\rm
Re}(T)\sim {\alpha^\prime}^2/R^4\ll 1$,  the magnetic flux
$h=\frac{1}{8\pi^2}\int F\wedge F$ might have a large value, which
would make $hS$ in $f_a^{(0)}$ non-negligible. The inclusion of $hS$
in $f_a^{(0)}$ suggests that the non-perturbative superpotential
should be generalized also as \bea W=w_0+\sum_i A_i
e^{-(a_iT_i+b_iS)}, \eea where $A_i={\cal O}(1)$, and $b_iS/a_iT_i$
might have a sizable value. For $S$ fixed at $\langle S\rangle=S_0$
by flux and a sequestered uplifting operator with ${\cal P}_0=$
constant, it is straightforward to repeat the previous analysis to
compute the vacuum values of $F^i$ and $T_i$ determined by the above
generalized form of moduli superpotential. One then finds
\cite{CJ}\bea a_iT_i+b_iS_0&=&\ln(M_{Pl}/m_{3/2})\left[1+{\cal
O}\left(\frac{1}{\ln(M_{Pl}/m_{3/2})}\right)\right], \nonumber \\
a_i F^i&=& 2m_{3/2}\left[1+{\cal
O}\left(\frac{1}{\ln(M_{Pl}/m_{3/2})}\right)\right],\eea
 which yields \bea
\left(\frac{M_a}{g_a^2}\right)_{\rm TeV}&=& \frac{1}{2}\sum_i
k_iF^i+\frac{1}{16\pi^2}b_am_{3/2}\nonumber \\
&=&
 \left(1+
\frac{\ln(M_{Pl}/m_{3/2})}{16\pi^2}g_{GUT}^2
b_a\alpha\right)\frac{M_0}{g_{GUT}^2}, \eea where \bea
\frac{1}{g_{GUT}^2}&=&\sum_ik_i{\rm Re}(T_i)+h{\rm
Re}(S_0),\nonumber \\
M_0&=&F^i\partial_i\ln({\rm
Re}(f_a^{(0)})=\frac{1}{2}g_{GUT}^2\sum_ik_iF^i,\nonumber
\\ \alpha&=&\frac{m_{3/2}}{M_0\ln(M_{Pl}/m_{3/2})}=\frac{h{\rm
Re}(S_0)+\sum_i k_i{\rm Re}(T_i)}{\sum_i [\frac{k_ib_i}{a_i}{\rm
Re}(S_0)+k_i{\rm Re}(T_i)]}.\eea Note that $\alpha\rightarrow 1$ in
the limit ${\rm Re}(S_0)/{\rm Re}(T_i)\rightarrow 0$, i.e. in the
limit when the magnetic flux-induced ${\cal O}(\alpha^{\prime 2})$
correction in $f_a^{(0)}$ is negligible. At any rate, the above
result gives rise to \bea M_1:M_2:M_3
\,\simeq\, (1+0.66\alpha):(2+0.2\alpha):(6-1.8\alpha),\eea where now
$\alpha$ can take a generic value of order unity.

\subsubsection{Partial KKLT stabilization \cite{PKKLT}}

In the partial-KKLT stabilization scenario, some K\"ahler moduli ($T_i$)
are stabilized by a non-perturbative superpotential, while the
remaining K\"ahler moduli ($X_p$)  are stabilized by the sequestered
uplifting potential. This scenario is an interesting generalization
of KKLT stabilization in which one combination of ${\rm Im}(X_p)$
can be identified as the QCD axion solving the strong CP problem
\cite{PKKLT}.

Like the case of KKLT stabilization, the SUSY-breaking brane is
assumed to be stabilized at the IR end of a warped throat, and thus
the resulting uplifting operator $\theta^2\bar{\theta}^2{\cal P}_0$
is independent of the entire set of K\"ahler moduli $T_I=(T_i,X_p)$:
 \bea K&=&K_0(T_i+T^*_i, X_p+X_p^*),
\nonumber \\
W&=&w_0+\sum_i A_i e^{-a_iT_i}, \nonumber \\
{\cal P}_0&=&\mbox{constant}. \eea It is also assumed that the model
allows a supersymmetric configuration satisfying \bea D_iW=0, \quad
D_pW=W\partial_pK=0.\eea One simple such example would be \bea
K&=&-2\ln[(\Phi_1+\Phi_1^*)^{3/2}-(\Phi_2+\Phi_2^*)^{3/2}-(\Phi_3+\Phi_3^*)^{3/2}],
\nonumber \\
W&=& w_0+A_1e^{-a_1\Phi_1}+A_2e^{-a_2(\Phi_2+\Phi_3)},\eea for which
one can rewrite the effective SUGRA in terms of $T_1=\Phi_1$,
$T_2=\Phi_2+\Phi_3$, and $X_1=\Phi_2-\Phi_3$. Although $X_p$ are
stabilized by the uplifting potential $V_{\rm lift}=e^{2K/3}{\cal
P}_0$, while $T_i$ are stabilized by non-perturbative
superpotential,  it turns out that the $F^I/(T_I+T_I^*)$ are again
{\it universal} for the entire K\"ahler moduli $T_I=(T_i,X_p)$ as
long as the moduli K\"ahler potential obeys the no-scale relation
\cite{PKKLT}: \bea \label{noscale1}
\sum_JK^{I\bar{J}}\partial_{\bar{J}}K=-(T_I+T_I^*), \eea which is
indeed satisfied at leading order in the $\alpha^\prime$-expansion.
As a result, the partial KKLT stabilization also gives rise to the
mirage pattern of gaugino masses as the KKLT stabilization, although
the details of moduli stabilization are quite different.

Similarly to the KKLT case, one can compute $F^I$ in two steps:
first start with the supersymmetric solution and then compute the
vacuum shift: \bea \delta T_I=(\delta T_i,\delta X_p)=\delta
\phi_I+i\delta a_I\eea caused by $V_{\rm lift}$. Again, for
$A_i={\cal O}(1)$, $D_iW=0$ in the first step determines $T_i$ as
\bea \label{susysol1} a_i T_i= \ln(M_{Pl}/m_{3/2})\left[1+{\cal
O}\left(\frac{1}{\ln(M_{Pl}/m_{3/2})}\right)\right],\eea regardless
of the detailed form of $K$.
 It is also obvious that $V_{\rm lift}$ does not induce any vacuum
shift in the axion direction: $\delta a_I=0$. As they are stabilized
by a non-perturbative superpotential, $T_i$ get a mass of ${\cal
O}(m_{3/2}\ln(M_{Pl}/m_{3/2}))$, and thus $\delta\phi_i={\cal
O}(1/[\ln(M_{Pl}/m_{3/2})]^2)$. On the other hand, since $W$ is
independent of $X_p$, ${\rm Re}(X_p)$ has a mass of ${\cal
O}(m_{3/2})$. One might then expect that $V_{\rm lift}$ causes a
large vacuum shift in the direction of ${\rm Re}(X_p)$, which would
result in a breakdown of our perturbative expansion.  However, this
is not the case since the supersymmetric configuration satisfying
$D_pW=W\partial_pK=0$ is a stationary point of $V_{\rm
lift}=e^{2K/3}{\cal P}_0$ for a constant ${\cal P}_0$. In fact,
$\delta\phi_p$ is induced by a K\"ahler mixing between ${\rm
Re}(X_p)$ and ${\rm Re}(T_i)$, and as a result $\delta\phi_p$ has
the same order of magnitude as $\delta\phi_i={\cal
O}(1/[\ln(M_{Pl}/m_{3/2})]^2)$.

Expanding $V_{\rm TOT}$ around the supersymmetric configuration
$\vec{T}_0$, we find \bea
 V_{\rm
TOT}&=&-3m_{3/2}^2(\vec{T}_0) +V_{\rm
lift}(\vec{T}_0)+2\partial_IV_{\rm lift}(\vec{T}_0)\delta\phi_I
+\left(m_\phi^2\right)_{IJ}\delta\phi_I\delta\phi_J +..., \eea where
the moduli mass matrix is given by\bea
 \left(m^2_\phi\right)_{ij} &=&
a_ia_jK^{i\bar{j}}\partial_iK\partial_{\bar{j}}Km_{3/2}^2
\left[1+{\cal
O}\left(\frac{1}{\ln(M_{Pl}/m_{3/2})}\right)\right],\nonumber \\
\left(m^2_\phi\right)_{p i} &=& 2m^2_{3/2}K_{p\bar{i}},\quad
\left(m^2_\phi\right)_{pq}\,=\, 2m^2_{3/2}K_{p\bar{q}}, \eea where
we have used $V_{\rm lift}(\vec{T}_0)\simeq 3m_{3/2}^2$. Minimizing
$V_{\rm TOT}$ determines $\delta\phi_I=(\delta\phi_i,\delta\phi_p)$
as \bea \label{shift1} &&\sum_j
a_ia_jK^{i\bar{j}}\partial_{\bar{j}}K\delta\phi_j = -2, \nonumber
\\
&&\sum_q K_{p\bar{q}}\delta\phi_q = -\sum_j K_{p
\bar{j}}\delta\phi_j, \eea which show that $\delta\phi_i$ and
$\delta\phi_p$ are all of the order of  $1/[\ln(M_{Pl}/m_{3/2})]^2$.
For such small vacuum shifts, the moduli $F$-components are obtained
to be \bea F^i&=&-e^{K/2}\sum_J
K^{i\bar{J}}(D_JW)^*\nonumber \\
&=&-m_{3/2}\sum_j a_jK^{i\bar{j}}\partial_{\bar{j}}K\delta
\phi_j\left[ 1+{\cal
O}\left(\frac{1}{\ln(M_{Pl}/m_{3/2})}\right)\right],\nonumber \\
F^p&=&-e^{K/2}\sum_JK^{p\bar{J}}(D_JW)^*\nonumber \\
&=& -m_{3/2}\sum_j a_j
K^{p\bar{j}}\partial_{\bar{j}}K\delta\phi_j\left[ 1+{\cal
O}\left(\frac{1}{\ln(M_{Pl}/m_{3/2})}\right)\right].\eea Then, using
(\ref{susysol1}) and (\ref{shift1}), we find \bea \label{F_result1}
&&\frac{F^i}{T_i+T_i^*}\,=\,\frac{m_{3/2}}{\ln(M_{Pl}/m_{3/2})}\equiv
M_0, \nonumber \\
&&\sum_q K_{p\bar{q}}F^{q*}\,=\,-\sum_jK_{p\bar{j}}F^{j*}, \eea

So far, we have {\it not} used any property  of the moduli K\"ahler
potential $K$. If $K$ obeys the no-scale condition (\ref{noscale1}),
the result on $F^p$ can be further simplified. Combining the
no-scale condition (\ref{noscale}) with $\partial_pK(\vec{T}_0)=0$,
one easily finds \bea \sum_j
(T_j+T_j^*)K_{p\bar{j}}=-\sum_q(X_q+X_q^*)K_{p\bar{q}}.\eea
Combining this with (\ref{F_result1}), one finds also \bea \sum_q
K_{p\bar{q}}\Big[F^q-(X_q+X_q^*)M_0\Big]=0,\eea which finally leads
to \bea
\frac{F^p}{X_p+X_p^*}&=&\frac{F^i}{T_i+T_i^*}\,=\,\frac{m_{3/2}}{\ln(M_{Pl}/m_{3/2})},
\nonumber \\
\frac{F^C}{C}&=& m_{3/2}+\frac{1}{3}F^I\partial_IK\,\simeq\,
m_{3/2}. \eea Thus, although the details of moduli stabilization in
the partial-KKLT scenario are quite different from the KKLT scenario, the
resulting pattern of SUSY breakdown is the same, and gives the
mirage pattern of low energy gaugino masses for the MSSM gauge
fields living on $D7$ branes.

\subsubsection{Uplifting via matter superpotentials \cite{LNR}}

The KKLT and partial KKLT scheme  assume that the uplifting of the
AdS vacuum is achieved by a brane-localized source of SUSY breaking
which is sequestered from the volume modulus and visible matter
fields. In Type IIB string theory, such sequestered SUSY breakdown
can be naturally realized if the SUSY-breaking brane is   stabilized
at the IR end of throat, while the visible sector and K\"ahler
moduli live on the bulk CY which corresponds to the UV end of
throat.

On the other hand, it has been pointed out that a successful
uplifting can be achieved also by a hidden sector realizing the
conventional spontaneous breakdown of $N=1$ SUGRA \cite{LNR,LLMNR}.
In such a scheme the hidden sector for uplifting
is not necessarily sequestered from the volume modulus and visible matter
fields.

More specifically, the model proposed in \cite{LNR} is
given by \bea
K&=&-3\ln(T+T^*)+ZZ^*, \nonumber \\
W&=&\phi(Z)+A(Z)e^{-aT}, \nonumber \\
f_a&=&T, \eea where $T$ is the volume modulus and  $Z$ is a hidden
matter field with nonzero $F^Z$  induced by a proper matter
superpotential $\phi(Z)$. Since the superpotential of $T$ has a
non-perturbative origin, there is still the natural possibility that
\bea |\,\partial_T^2 W|\gg |\,W|,\eea realizing \bea m_T \gg
m_{3/2}, \quad \frac{F^T}{T+T^*}\ll m_{3/2}. \eea
On the other
hand, the condition of nearly vanishing cosmological constant
requires \cite{LLMNR} \bea F^Z={\cal O}(m_{3/2}), \quad
\frac{F^C}{C}={\cal O}(m_{3/2}).\eea In such a case, the conformal
anomaly mediation from $F^C/C={\cal O}(m_{3/2})$ and also the
one-loop
 contributions from $F^Z={\cal O}(m_{3/2})$ can become
equally important to the tree level contribution to $M_a/g_a^2$
from $|F^T|\ll m_{3/2}$. In particular, string thresholds might
depend on $Z$, and thus give a  UV sensitive contribution to
$M_a/g_a^2$ through the term $\frac{1}{8\pi^2}F^Z\partial_Z\Omega_a$
in (\ref{lowscaleratio}), which could  spoil the
predictability of the scheme.

However, this potential difficulty can be  avoided by making a
simple assumption on the property of the hidden matter $Z$, for
instance by assuming a discrete symmetry  $Z\rightarrow -Z$ which is
broken only by non-perturbative dynamics that is
responsible for SUSY breakdown. This
(approximate) symmetry ensures that the vacuum values of
$\partial_Z\ln(e^{-K_0/3}Z_m)$ and $\partial_Z\Omega_a=0$ are
negligibly small, and therefore the one-loop contribution to $M_a/g_a^2$
is dominated by the conformal anomaly mediation.
Note that  the
above model does not necessarily take a sequestered form:
$e^{-K/3}=\Omega_1(T,T^*)+\Omega_2(Z,Z^*)$ and $W=W_1(T)+W_2(Z)$.
This has several interesting consequences. First of all, it allows
more freedom for the relative importance of anomaly mediation, i.e.
a much larger range of values of $\alpha$ is allowed. Secondly,
while a mirage pattern is expected for the gaugino masses, this is
not necessarily true for the squark and slepton masses as well
\cite{LNR,LLMNR}.

To see the pattern of low energy gaugino masses obtainiable by the matter
uplifting scenario, let us consider an example with a rather large
value of $\alpha$. It is the example of ref. \cite{LNR} with a
superpotential  $W$ expanded around the vacuum configuration $\langle
T\rangle= 2$ and $\langle Z\rangle =0$: \bea
W&=&\epsilon\left[\,0.577+Z+0.441(T-2)+0.592 Z^2+9.595(T-2)^2\right.
\nonumber \\
&& \left. +\,0.114 Z^3
+0.220Z^2(T-2)+46.451(T-2)^3+\mbox{higher order terms}\,\right],\eea
where $\epsilon$ is a small parameter of ${\cal O}(m_{3/2})$ which
might be generated by nonperturbative dynamics, like e.g. gaugino
condensation. This particular example of $W$ stabilizes $T$ and $Z$
at a nearly Minkowski vacuum with \bea F^Z\,=\,\frac{\epsilon}{8},
\quad {F^T}\,=\, \frac{\epsilon}{200},\quad
m_{3/2}=\frac{\epsilon}{14}. \eea The resulting $M_a/g_a^2$ take a
mirage pattern: \bea \left(\frac{M_a}{g_a^2}\right)_{\rm TeV}&=&
\frac{1}{2}F^T+\frac{b_a}{16\pi^2}m_{3/2} \nonumber \\
&=&\left(1+\frac{\ln(M_{Pl}/m_{3/2})}{16\pi^2}g_{GUT}^2b_a\alpha\right)\frac{M_0}{g_{GUT}^2},\eea
with \bea \alpha\simeq 1.7,\eea and thus \bea M_1:M_2:M_3\simeq 1:
1.1: 1.4. \eea

Although the above specific  example of $W$ leads to the mirage
parameter $\alpha\simeq 1.7$, one can construct different
superpotentials giving a different value of $\alpha$ within the range
of order unity and smaller. This is in contrast to the
KKLT and partial KKLT schemes with minimal gauge kinetic
function $f_a=T$, where we obtain $\alpha\simeq 1$.
This difference is due to the fact that the SUSY breaking
sector is sequestered from $T$ in KKLT and partial KKLT, while it is
not in the matter uplifting scenario. As we have noticed, $\alpha\neq 1$
can be obtained also in KKLT and partial KKLT while keeping the SUSY
breaking sector sequestered from $T$ if the magnetic flux-induced
corrections to $f_a$ are sizable. Note also that a non-sequestered
coupling between $Z$ and the visible matter fields can give a
contribution of ${\cal O}(m_{3/2}^2)$ to the sfermion mass-squared
in the matter uplifting scenario \cite{LNR}.

\subsubsection{Deflected anomaly mediation \cite{DAM}}

Deflected anomaly mediation \cite{DAM} has been proposed as a
solution to the tachyonic slepton problem of the original anomaly
mediation scenario \cite{anomalymediation}. The scheme assumes a
gauge-charged messenger sector  which experiences a non-decoupled
SUSY breakdown triggered by the $F$-component of the SUGRA
compensator. Such a gauge messenger sector then provides a
gauge-mediated contribution to soft masses  comparable to anomaly
mediation at the threshold scale of gauge-charged messenger fields.
This deflects the soft masses  from the anomaly mediation trajectory
at scales below the gauge threshold scale, thereby avoiding
tachyonic sleptons.

 A simple example of such model is provided by the following messenger
sector superpotential containing a scale {\it non-invariant}
term:\bea W_{\rm mess}=\lambda_\Phi X_\Phi
\Phi^c\Phi+M_*^{3-n}X_\Phi^n,\eea where $n\neq 3$, $M_*$ is a
model-dependent mass parameter, $X_\Phi$ is a singlet chiral
superfield, and finally the gauge-charged $\Phi+\Phi^c$ are assumed
to form a GUT representation to maintain the gauge coupling
unification at $M_{GUT} \sim 2\times 10^{16}$ GeV. The chiral
compensator $C$ couples to $X_\Phi$ at tree level through the scale
non-invariant term $M_*^{3-n}X_\Phi^n$ ($n\neq 3$), thereby fixing
the vacuum value of $X_\Phi$. One then finds that $X_\Phi$ is
stabilized at $\langle X_\Phi\rangle \gg m_{3/2}$ for $M_*\gg
m_{3/2}$ if $n<0$ or $n>3$ \cite{DAM}, explicitly \bea \langle
X_\Phi\rangle=M_\Phi+\theta^2F^{X_\Phi},\eea where \bea M_\Phi&\sim&
m_{3/2}^{1/(n-2)}M_*^{(n-3)/(n-2)}, \nonumber \\
\frac{F^{X_\Phi}}{M_\phi}&=& -\frac{2}{n-1}\frac{F^C}{C}.\eea In
fact, even in the absence of the scale non-invariant term
$M_*^{n-3}X_\Phi^n$ ($n>3$ or $n<0$) in the superpotential, $X_\Phi$
still can be stabilized by radiative corrections to its K\"ahler
potential, yielding \cite{DAM}\bea
\frac{F^{X_\Phi}}{M_\phi}\,\simeq\, -\frac{F^C}{C}.\eea

As $F^{X_\Phi}/M_\Phi={\cal O}(F^C/C)$ in the above case, the
resulting gauge-mediated contribution to soft masses is comparable
to the anomaly-mediated ones. Applying the result of Eq.
(\ref{lowscaleratio}) to this case, one easily finds
  \bea \left(\frac{M_a}{g_a^2}\right)_{\rm
TeV}&=&\frac{b_a}{16\pi^2}\frac{F^C}{C}-\frac{1}{8\pi^2}\sum_\Phi
C^\Phi_a\frac{F^{X_\Phi}}{M_\Phi} \nonumber \\
&=&
\frac{1}{16\pi^2}\frac{F^C}{C}\left(\,b_a+\frac{2}{n-1}N_\Phi\,\right),\eea
where $n\geq 3$ or $n<0$, and $N_\Phi=\sum_\Phi
(C_a^\Phi+C_a^{\Phi^c})$ is the number of messenger pairs
$\Phi+\Phi^c$. This is the mirage pattern of gaugino masses with
\bea
\alpha=\frac{16\pi^2}{g_{GUT}^2\ln(M_{Pl}/m_{3/2})}\frac{n-1}{2N_\Phi}.\eea
We note that, for a given value of $\alpha$, the sfermion spectrum
in deflected anomaly mediation \cite{DAM} takes a different pattern
than that of mirage mediation \cite{mirage2}, although the gaugino
masses share the same mirage pattern.


\subsection{Anomaly pattern}

In anomaly mediation \cite{anomalymediation}, SUSY breaking fields
$X^I$ are assumed to be (effectively) sequestered from the visible
sector fields, which means that $F^I\partial_If^{(0)}_a$,
$\frac{1}{8\pi^2}F^I\partial_I(e^{-K_0/3}Z_i)$ and
$\frac{1}{8\pi^2}F^I\partial_I\Omega_a$  in Eq.(\ref{lowscaleratio})
are all subdominant compared to the SUGRA compensator mediated
contribution \cite{sequestering,conformalsequestering}. Then
$M_a/g_a^2$ are determined  as \bea
\left(\frac{M_a}{g_a^2}\right)_{\rm
TeV}\,\simeq\,\tilde{M}_a^{(1)}|_{\rm
conformal}=\frac{b_a}{16\pi^2}\frac{F^C}{C}. \eea If the effective
theory around TeV  is given by the MSSM, $b_a=(\frac{33}{5},1,-3)$
and the low energy gaugino masses take the anomaly pattern: \bea
\label{anomalypattern1} M_1:M_2:M_3 \,\simeq\, 3.3:1:9.\eea One
example of string-based scenario which can give  the anomaly pattern
of gaugino masses is the fluxed compactification of Type IIB string
theory with visible sector living on $D3$ branes \cite{mirage1}.

\subsubsection{Visible sector on $D3$ branes in Type IIB string theory}

In fluxed Type IIB compactification, the ISD 3-form flux fixes the
dilaton $S$ and complex structure moduli $U$ at a supersymmetric
solution of $D_SW_{\rm flux}=D_UW_{\rm flux}=0$. As the flux-induced
masses of $S$ and $U$ are hierarchically heavier than $m_{3/2}$, the
$F$-components of $S$ and $U$ remain to be negligible even after the
subsequent SUSY-breaking effect is taken into account \footnote{In
fact, this is not a precise statement for the KKLT scenario. In the KKLT
scenario, the complex structure modulus
 describing the collapsing 3-cycle of the warped throat can develop
a large $F$-component through its direct coupling to the SUSY
breaking brane at the IR end of throat. However, such particular
complex structure modulus is sequestered from the visible sector
$D3$ branes which are assumed to be stabilized somewhere in the bulk
CY, and thus does not affect the soft terms of visible fields.}:\bea
|F^{S,U}|\,\sim\, \frac{m_{3/2}^2}{m_{S,U}}\,\ll\,
\frac{m_{3/2}}{8\pi^2}, \eea while the $F$-components of K\"ahler
moduli $T_i$ can be ${\cal O}(m_{3/2}/8\pi^2)$ or bigger, depending
upon how $T_i$ are stabilized.

To be specific, let us first consider the KKLT-type stabilization of
$T_i$, yielding the following pattern of $F$-terms \cite{mirage1}:
\bea \frac{F^i}{T_i+T_i^*}={\cal
O}\left(\frac{m_{3/2}}{8\pi^2}\right), \quad \frac{F^C}{C}\simeq
m_{3/2}.\eea
 For a gauge field on $D3$,
the corresponding gauge kinetic function is given by \bea
f_{3a}=S\eea at leading order in the $\alpha^\prime$ and string loop
expansions. The only possible correction to $f_{3a}$ allowed by the
axionic shift symmetries takes the form: $\Delta
f_{3a}=\epsilon_aT$, where $\epsilon_a$ are real (discrete)
constants. However any nonzero value of $\epsilon_a=0$ can not give
a sensible behavior of $D3$ brane gauge coupling in the large volume
limit ${\rm Re}(T)/{\rm Re}(S) \gg 1$\footnote{Note that ${\rm
Re}(S)\propto 1/g_{st}$ and ${\rm Re}(T)\propto R^4/g_{st}$, where
$g_{st}$ and $R$ denote the string coupling and the compactification
radius, respectively.}.
 Thus the above form of $D3$
gauge kinetic function is exact up to non-perturbative effects of
${\cal O}(e^{-8\pi^2 S})$ or ${\cal O}( e^{-8\pi^2 T})$. Applying
the  $F$-components in KKLT scenario to $M_a/g_a^2$ in
(\ref{lowscaleratio}) with $f_a=S$, one easily finds that
$M_a/g_a^2$ are dominated by the conformal anomaly contribution
$b_aF^C/16\pi^2C={\cal O}(m_{3/2}/8\pi^2)$, and thereby the
resulting low energy gaugino masses take the anomaly pattern
(\ref{anomalypattern1}). Note that the contributions from
$\frac{1}{8\pi^2}F^i\partial_i\ln(e^{-K_0/3}Z_m)$ and
$\frac{1}{8\pi^2}F^i\partial_i\Omega_a$ are of the order of
$m_{3/2}/(8\pi^2)^2$, {\it independently  of} how the matter
K\"ahler metric $Z_m$ and the string threshold correction $\Omega_a$
depend on $T_i$.

In KKLT scenario, $F^T/(T+T^*)={\cal O}(m_{3/2}/8\pi^2)$ and
$F^C/C\simeq m_{3/2}$ guarantees that $M_a/g_a^2$ are dominated by
the conformal anomaly mediation irrespective of the $T$-dependence
of $e^{-K_0/3}Z_m$ and $\Omega_a$. However, in flux-induced
SUSY-breaking scenario with $T$ stabilized by the perturbative
K\"ahler corrections in (\ref{corrected_Kahler}), we have an
opposite hierarchy of $F$-terms \cite{choi}: \bea
\frac{F^T}{T+T^*}\simeq m_{3/2},\quad \frac{F^C}{C}\,\simeq\,
\frac{0.5\xi_1}{(T+T^*)^{3/2}} m_{3/2}\,\ll\, m_{3/2}. \eea As a
result, one needs a higher degree of sequestering, i.e. a strong
suppression of $\partial_T\ln(e^{-K_0/3}Z_m)$ and
$\partial_T\Omega_a$, in order to achieve a conformal  anomaly
domination in flux-induced SUSY breaking scenario. In fact, we have
$\partial_T\ln(e^{-K_0/3}Z_m)=0$ for the matter K\"ahler metric
$Z_m$ on $D3$ at leading order in the $\alpha^\prime$ and string
loop expansions. It is also expected that the tree-level masses of
open string modes on $D3$ are determined by the local physics on the
$D3$ world volume, thus are independent of the volume modulus $T$.
This would result in $\partial_T\Omega_a=0$ for one-loop string
threshold corrections. These  indicate that the anomaly pattern of
$D3$ gaugino masses might be possible in flux-induced SUSY breaking
scenario also, although it requires that the sequestering persists
for higher order matter K\"ahler metric and string loop threshold
corrections.

As is well known,  pure anomaly mediation for sfermion masses is
problematic, as it leads to tachyonic sleptons at the TeV scale.
However, it might be possible to avoid tachyonic sleptons while
keeping the anomaly pattern of gaugino masses maintained
\cite{dterm_anomaly}, for instance by introducing a $D$-term
contribution to sfermion masses.


\subsection{Schemes with strong model dependence}

In SUSY-breaking scenarii leading to the above  three patterns
of gaugino masses, the UV sensitive string and/or KK threshold
contributions $\tilde{M}_a^{(1)}|_{\rm string}$ to the ratios
$M_a/g_a^2$ are always subleading effects compared to the UV
insensitive contributions calculable within the 4D effective theory.
This is the reason that we can make
reliable predictions for the low energy gaugino
mass ratios. However, in certain cases, $\tilde{M}_a^{(1)}|_{\rm
string}$ becomes one of the dominant contributions, and then we
loose the predictive power as the low energy gaugino masses become
sensitive to the UV physics above the 4D cutoff scale $\Lambda$.
Here we present two such examples; one is the volume
moduli-dominated SUSY breakdown in perturbative heterotic string
theory \cite{kaplunovskylouis,binetruy} and the other is the $M$
theory compactification on $G_2$ manifolds with moduli stabilization
by non-perturbative dynamics \cite{acharya,acharya1}.

\subsubsection{Moduli dominated scenario in heterotic orbifold
compactification}

As an example of the scheme in which $\tilde{M}_a^{(1)}|_{\rm
string}$ is one of the dominant contributions to $M_a/g_a^2$, let us
consider  the heterotic string compactification on orbifolds for
which the tree-level K\"ahler potential and gauge kinetic function
are given by \bea K&=&-\ln(S+S^*)-\sum_i\ln(T_i+T_i^*)+
\prod_i(T_i+T_i^*)^{n_i^m}Q^*_mQ_m, \nonumber \\
f_a^{(0)}&=& S,\eea where $T_i$ ($i=1,2,3$) are the volume moduli of
the underlying toroidal compactification. Using the constraint from
the $SL(2,Z)$ modular invariance for each $T_i$, the one-loop
gaugino masses including string threshold correction  have been
obtained in \cite{binetruy}. Applying the result of \cite{binetruy}
for the following pattern of  volume moduli-dominated SUSY
breakdown: \bea \left|\frac{F^S}{S+S^*}\right|\,\lesssim\, {\cal
O}\left(\frac{m_{3/2}}{8\pi^2}\right), \quad
\frac{F^i}{T_i+T_i^*}={\cal O}(m_{3/2}),\eea and ignoring the
subdominant pieces, we find \bea
\label{unpredictable}\left(\frac{M_a}{g_a^2}\right)_{\rm TeV}&=&
\tilde{M}_a^{(0)}+ \tilde{M}_a^{(1)}|_{\rm
conformal}+\tilde{M}_a^{(1)}|_{\rm Konishi}+\tilde{M}_a^{(1)}|_{\rm
string} \eea
 where \bea 
\tilde{M}_a^{(0)}&=&\frac{1}{2}F^I\partial_If_a^{(0)}=\frac{1}{2}F^S, \nonumber \\
\tilde{M}_a^{(1)}|_{\rm conformal}
&=&\frac{1}{16\pi^2}b_a\frac{F^C}{C}
\nonumber \\
&=&-\frac{1}{16\pi^2}(3C_a-\sum_mC_a^m)\left(m_{3/2}-\frac{1}{3}\sum_i\frac{F^i}{T_i+T_i^*}\right) \nonumber \\
\tilde{M}_a^{(1)}|_{\rm
Konishi}&=&-\frac{1}{8\pi^2}\sum_mC_a^mF^I\partial_I\ln(e^{-K_0/3}Z_m)
\nonumber
\\
&=&-\frac{1}{8\pi^2}\sum_{i,m}C_a^m\left(\frac{1}{3}+n_i^m\right)\frac{F^i}{T_i+T_i^*},
\nonumber \\
\tilde{M}_a^{(1)}|_{\rm string}&=&\frac{1}{16\pi^2}\sum_iF^i\left[\,
\frac{1}{3}\delta_{GS}\partial_i\ln\Big((T_i+T_i^*)\eta^2(T_i)\Big)
\right.\nonumber
\\
&&+\,\left.\Big(C_a-\sum_mC_a^m(1+2n_i^m)\Big)\partial_i\ln\eta^2(T_i)\,\right],
\eea where $\eta(T)=e^{-\pi T/12}\prod_{n=1}^{\infty}(1-e^{-2\pi
nT})$ is the Dedekind function, and $\delta_{GS}$ is the coefficient
of the Green-Schwarz counterterm.
Rearranging the above result, we finally find  \bea
\left(\frac{M_a}{g_a^2}\right)_{\rm TeV}
&=&\frac{1}{2}F^S+\frac{b_a}{16\pi^2}m_{3/2} \nonumber
\\
&+&\frac{1}{16\pi^2} \sum_i
\left(\frac{F^i}{T_i+T_i^*}\right)\kappa(T_i)
\left(\frac{\delta_{GS}}{3}+C_a-\sum_m (1+2n_i^m)C_a^m \right),\eea
where \bea \kappa(T_i)=1+(T_i+T_i^*)\partial_i\ln\eta^2(T_i). \eea

The above result shows that the low energy gaugino masses in the
volume moduli-dominated SUSY-breaking scenario in heterotic string
theory are indeed quite sensitive to  the string threshold
$\tilde{M}_a^{(1)}|_{\rm string}$. Although, in this case, one could
determine $\tilde{M}_a^{(1)}|_{\rm string}$ using the $SL(2,Z)$
modular invariance \cite{binetruy}, usually this is not  possible in
more generic compactifications, and then we loose the predictive
power.

The volume moduli domination in heterotic string might be achieved
in the racetrack scenario of dilaton stabilization \cite{racetrack}.
However, the racetrack stabilization typically leads to an AdS
vacuum and still lacks a mechanism lifting this AdS vacuum to a dS
vacuum while keeping the volume-moduli domination.

We note that the gaugino mass pattern (\ref{unpredictable}),
although generically not predictive since it depends on many model
parameters, corresponds to the mirage pattern in a special
circumstance in which both $T_i$ and $F^i$ have {\it universal}
vacuum values,\bea T_1&=&T_2\,=\,T_3\,=\,T, \nonumber \\
 F^{T_1}&=&F^{T_2}\,=\,F^{T_3}\,=\,F^T,\eea
 and also all
gauge charged matter fields originate from the untwisted sector, so
that \bea \sum_i n_i^m=-1.\eea In such case, (\ref{unpredictable})
gives \bea \left(\frac{M_a}{g_a^2}\right)_{\rm TeV}&=&
\frac{1}{2}F^S
+\frac{1}{16\pi^2}\kappa(T)\delta_{GS}\frac{F^T}{T+T^*}
+\frac{b_a}{16\pi^2}\left(m_{3/2}-\kappa(T)\frac{F^T}{T+T^*}\right),
\eea leading to \bea M_1:M_2:M_3\simeq
(1+0.66\alpha):(2+0.2\alpha):(6-1.8\alpha)\eea with \bea
\alpha=\left(\frac{16\pi^2}{g_{GUT}^2\ln(M_{Pl}/m_{3/2})}\right)\left(
\frac{(T+T^*)m_{3/2}-\kappa(T)F^T}{8\pi^2(T+T^*)F^S+\kappa(T)\delta_{GS}F^T}\right).\eea

\subsubsection{$M$ theory on $G_2$ manifolds}

Another scheme in which the string thresholds  can significantly
affect the  gaugino mass ratios is the recently studied $M$ theory
compactification on $G_2$ manifolds  in which the moduli are
stabilized by non-perturbative dynamics \cite{acharya,acharya1}. The
K\"ahler potential and gauge kinetic functions of 4D effective
theory are given by \bea K&=&-\sum_i
n_i\ln(T_i+T_i^*)+Z_\phi(T_i+T_i^*)\phi\phi^*
\nonumber \\
&&+\,Z_m(T_i+T_i^*,\phi,\phi^*)Q^*_mQ_m+...\nonumber \\
f_a&=&\sum_i k_{i}T_i,\eea where
 $n_i$ are positive rational
numbers satisfying $\sum_in_i=7/3$, $k_{i}$ are integers, and the
ellipsis stands for the terms higher order in $\phi$ and $Q_m$. Here
${\rm Re}(T_i)$ correspond to the 3-cycle volume moduli, $\phi$ is a
composite  hidden matter whose $F$-component is crucial for the
model to have a phenomenologically viable dS or Minkowski vacuum,
and $Q_m$ are the visible matter superfields. Under the assumption
of non-perturbative dynamics generating for instance a
superpotential of the (racetrack) form: \bea
W=A_1\phi^ke^{-a_1f_h}+A_2e^{-a_2f_h} \eea for a hidden sector gauge
kinetic function $f_h=\sum_i \tilde{k}_{i}T_i$, it has been noticed
that $\phi$ and $T_i$ can be stabilized (except for some axion
components which are harmless) at a SUSY-breaking dS vacuum with the
following  features \cite{acharya1}: \bea &&\phi={\cal O}(1),\quad
\frac{F^\phi}{\phi}={\cal O}(m_{3/2}),\quad \frac{F^C}{C}={\cal
O}(m_{3/2}),\nonumber \\
&& \frac{F^i}{T_i+T_i^*}={\cal
O}\left(\frac{m_{3/2}}{\ln(M_{Pl}/m_{3/2})}\right).\eea Note that
$F^\phi={\cal O}(m_{3/2})$ is crucial in order for the vacuum energy
density $\langle V\rangle=K_{I\bar{J}}F^IF^{J*}-3|m_{3/2}|^2$ to be
nearly vanishing.

With the above pattern of SUSY-breaking $F$-components, the one-loop
contributions to $M_a/g_a^2$  are generically comparable to the
tree-level contribution, thus should be carefully taken into
account.
Explicitly, we have \bea \left(\frac{M_a}{g_a^2}\right)_{\rm TeV}&=&
\tilde{M}_a^{(0)}+ \tilde{M}_a^{(1)}|_{\rm
conformal}+\tilde{M}_a^{(1)}|_{\rm Konishi}+\tilde{M}_a^{(1)}|_{\rm
string}, \eea where \bea \tilde{M}_a^{(0)}&=&\frac{1}{2}\sum_i
k_iF^i={\cal O}\left(\frac{m_{3/2}}{\ln(M_{Pl}/m_{3/2})}\right),
\nonumber \\
\tilde{M}_a^{(1)}|_{\rm conformal}&=&
\frac{1}{16\pi^2}b_a\left(m_{3/2}+\frac{1}{3}F^\phi\partial_\phi
K\right),
\nonumber \\
\tilde{M}_a^{(1)}|_{\rm Konishi}&=&-\frac{1}{8\pi^2}\sum_m
C_a^mF^\phi\partial_\phi\ln(e^{-K_0/3}Z_m),\nonumber\\
\tilde{M}_a^{(1)}|_{\rm string}&=&
\frac{1}{8\pi^2}F^\phi\partial_\phi\Omega_a,\eea where
 $K_0$
is the K\"ahler potential of $T_i$ and $\phi$,
$\frac{1}{8\pi^2}\Omega_a$ are the M-theory thresholds for the
visible gauge coupling superfields, and we have ignored the
subleading part of ${\cal O}(F^i/8\pi^2)$.
 As for the relative importance of each  of the
one-loop contributions, one can consider two distinct possibilities.
The first possibility is that the hidden matter $\phi$ is
sequestered from the visible sector, which would mean \bea
\Gamma_\phi\equiv\partial_\phi\ln(e^{-K_0/3}Z_m) \simeq 0,\qquad
\Omega_{a\phi}\equiv\partial_\phi \Omega_a\simeq 0.\eea This is in
principle an open possibility as $\phi$ lives at a point-like
conical singularity spatially separated from the 3 cycle (and the
conical singularities in it) of the visible gauge fields
\cite{acharya1}. In such sequestered case, $M_a/g_a^2$ are
determined by the universal $\tilde{M}_a^{(0)}$ and the comparable
 conformal anomaly contribution $\tilde{M}_a^{(1)}|_{\rm
conformal}$, and thus take the mirage pattern.

 However, sequestering is not a generic consequence of
geometric separation, but arises only in special circumstances such
as the case of 5D bulk geometry or a geometric separation by warped
throat \cite{nonsequestering}. In case that $\phi$ is not
sequestered from the visible sector, which is actually the case
assumed in \cite{acharya,acharya1}, both $\Gamma_\phi$ and
$\Omega_{a\phi}$ are expected to be of order unity. If one assumes
(as is done explicitely in \cite{acharya1}) that the hidden matter
K\"ahler metric takes the minimal form and the visible matter
K\"ahler metrics are independent of $\phi$, i.e. \bea
K=-\sum_in_i\ln(T_i+T_i^*)+\phi\phi^*+Z_m(T_i+T_i^*)Q_mQ_m^*,\eea
the resulting value of $\Gamma_\phi$ is indeed of order unity: \bea
\Gamma_\phi=-\frac{1}{3}\phi^*={\cal O}(1). \eea Although the order
of magnitude is unchanged, including the terms higher order in
$\phi$ can significantly change the size of $\Gamma_\phi$, and thus
the size of the Konishi anomaly contribution
$\tilde{M}_a^{(1)}|_{\rm Konishi}$. Furthermore, the masses of
superheavy gauge-charged $M$-theory matter fields $Q_H+Q_H^c$
\cite{mthreshold} living at the same conical singularity as $Q_m$
can have a unsuppressed $\phi$-dependence through for instance
$e^{-K_0/3}Z_H$ where $Z_H$ denotes the K\"ahler metric of $Q_H$ or
through the higher-dimensional superpotential coupling between
$Q_H+Q_H^c$ and the hidden matter $\phi$ such as $\phi^lQ_HQ_H^c$.
Since $\phi$ has a vacuum value of order unity in the unit with
$M_{Pl}=1$, this eventually yields a sizable$M$-theory threshold to
gaugino masses: \bea \Omega_{a\phi}={\cal O}(1).\eea In this case,
the gaugino mass ratios can be determined only when one can reliably
compute the values of the highly UV sensitive $\Gamma_\phi$ and
$\Omega_{a\phi}$, which is not available with our present
understanding of $M$ theory compactification.


\section{Summary and Search strategy}

In the search for supersymmetry at the LHC,
the identification of gauginos will play a crucial role.
Predictions for the masses of gauginos are rather robust
and seem to favor few distinctive patterns. Of those, the
mSUGRA pattern
\bea
\mbox{mSUGRA pattern:}\quad M_1:M_2:M_3\,\simeq\,
1:2:6 \eea
is shared by many schemes, such as
\begin{itemize}

\item gravity mediation \cite{gravitymediation},

\item various schemes of dilaton/moduli mediation in string and M-theory including the flux-induced
SUSY breakdown \cite{kaplunovskylouis,nilles,ibanez},

\item gaugino mediation \cite{gauginomediation},

\item gauge mediation \cite{gaugemediation},

\item large volume compactification in Type IIB string theory \cite{quevedo}.

\end{itemize}
The mSUGRA pattern arises if the ratios $M_a/g_a^2$ in
(\ref{lowscaleratio}) are dominated by  universal tree level
contribution $\tilde{M}_a^{(0)}$ or by universal gauge threshold
contribution $\tilde{M}_a^{(1)}|_{\rm gauge}$. This pattern is
closely related to the gauge coupling constants in the TeV range and
rather independent of the ultraviolet properties of the underlying
scheme. It can appear independently of gauge coupling unification at
a large scale, although in some cases (like gauge mediation) such an
assumption seems to be required.

In the anomaly pattern
  \bea
\mbox{Anomaly pattern:}\quad  M_1:M_2:M_3\,\simeq\, 3.3:1:9, \eea
$M_a/g_a^2$ are dominated by the conformal anomaly contribution
$\tilde{M}_a^{(1)}|_{\rm conformal}$ related to the $SU(3)\times
SU(2)\times U(1)$ $\beta$-functions \cite{anomalymediation}. Schemes
that realize this pattern need a very strict separation of the
hidden SUSY-breakdown sector from the visible sector of the
supersymmetric standard model (MSSM), i.e. a strong sequestering
\cite{sequestering,conformalsequestering}. Because of this
restriction its appearance is rather rare and delicate. One
possibility can be found in Type IIB string theory with visible
sector  on D3 branes. In its pure form, anomaly mediation is
problematic, as it predicts tachyonic sleptons. This problem has to
be removed without disturbing the gaugino mass pattern.

The mirage pattern for gaugino masses
 \bea
\mbox{Mirage pattern:}\quad M_1:M_2:M_3
\,\simeq\, (1+0.66\alpha):(2+0.2\alpha):(6-1.8\alpha) \eea with
$\alpha={\cal O}(1)$ arises if $M_a/g_a^2$ are dominated by
$\tilde{M}_a^{(0)}$ and $\tilde{M}_a^{(1)}|_{\rm conformal}$ which
are comparable to each other. Schemes yielding the mirage pattern
have recently been identified in various versions of string and M
theory \cite{mirage1,mirage2}:
\begin{itemize}

\item KKLT moduli stabilization \cite{KKLT} in Type IIB string theory with visible sector on D7 branes,

\item partial KKLT moduli stabilization \cite{PKKLT},

\item uplifting via matter superpotentials \cite{LNR},

\item deflected anomaly mediation \cite{DAM}.

\end{itemize}
In these schemes, the leading contribution of moduli mediation is
suppressed by a factor $\log(M_{\rm Planck}/m_{3/2})$ such that the
contribution of the conformal anomaly mediation (suppressed by a
loop factor) becomes competitive, while the other (UV sensitive)
one-loop contributions, i.e. the Konishi anomaly contribution
$\tilde{M}_a^{(1)}|_{\rm Konishi}$ and the string threshold
correction $\tilde{M}_a^{(1)}|_{\rm string}$ in the formulae
(\ref{lowscaleratio}) and (\ref{components}),
 remain to be subleading.   In such a scheme
of mixed modulus-anomaly mediation the predictions for gaugino
masses are again pretty robust and reliable, while the patterns of
squark and slepton masses show stronger model dependence.

Besides the schemes leading to the above three patterns of gaugino
masses, one can imagine other scenario in which
$\tilde{M}_a^{(1)}|_{\rm Konishi}$ and/or $\tilde{M}_a^{(1)}|_{\rm
string}$ become important, thus give a different gaugino mass
pattern. If $\tilde{M}_a^{(1)}|_{\rm Konishi}$ from light matter
fields is important, $\tilde{M}_a^{(1)}|_{\rm string}$ from heavy
string or $M$ theory modes is expected to become important also.
We then loose the predictive power as the low energy gaugino masses
become sensitive to the UV physics above the 4D cutoff scale
$\Lambda$.
Examples of such scheme include the volume moduli-dominated SUSY
breakdown in perturbative heterotic string theory \cite{binetruy}
and the $M$ theory compactification on $G_2$ manifolds with moduli
stabilization by non-perturbative dynamics \cite{acharya,acharya1}.

Of course, the mass patterns identified so far correspond to the
parameters $M_1$, $M_2$ and $M_3$ of the MSSM and not yet to the
mass eigenstates. The challenge for phenomenological analyses will
be the connection of the $M_a$ ($a=1,2,3$) to the physical masses.
Sample spectra for the mirage pattern have been worked out in
\cite{mirage3}. The gluino mass is directly related to $M_3$,
whereas the LSP-neutralino can be a mixture of bino, wino as well as
Higgsino. The mSUGRA pattern would favor a bino-LSP with a ratio 1/6
compared to the gluino while in the anomaly mediation scheme we
might have a wino-LSP with ratio 1/9 to the gluino mass. Of course,
we expect mixed states and have to take into account a possibly
sizable Higgsino component. If the gluino to LSP ratio turns out to
be anomalously large, this could then be a signal for a
Higgsino-LSP. On the other hand, a small ratio of the gluino to the
LSP-mass (less than 6) might be a hint towards the mirage pattern.
Thus even with the knowledge of two of the $M_a$ parameters we might
already distinguish between the various different patterns.

If we have identified the LSP, most probably we shall also have some
information on the heavier neutralinos as the LSP might be the
end-product of a cascade decay. This would allow us to formulate sum
rules including all three $M_a$  to further check the patterns. Note
that the combination \bea \label{ratio} r= {\frac{1} {M_3}}\left(
2(M_1 + M_2) - M_3\right) \eea will approximately vanish both in the
mSUGRA and anomaly pattern. A nonvanishing $r$ would thus be a sign
of the mirage scheme and allow a determination of the mirage
mediation parameter $\alpha$. For the benchmark scenario with
$\alpha=1$ (intermediate mirage messenger scale) discussed earlier,
one would obtain $r\approx 0.84$, while for the case with $\alpha=2$
(TeV mirage scale) we get $r\approx 2.93$.

Having determined the pattern of gaugino masses we would then
include information on squark and slepton masses to break
possible degeneracies. Unfortunately, sfermion masses show a
stronger model dependence. Still, as we have seen in our
discussion in chapter 3, useful information could be extracted
once the pattern has been identified via the gaugino masses.
Such a discussion is beyond the scope of this paper and will
be the subject of future investigations\footnote{For some recent
discussion of this aspect, see ref. \cite{cohen}.}.
Still we think that
the values of the gaugino masses will give us the first hint
to unravel the underlying structure of supersymmetry breakdown.

\vspace{0.2cm} \vspace{5mm} \noindent{\large\bf Acknowledgments}
\vspace{5mm}

This work was partially supported by the European Union 6th
framework program MRTN-CT-2004-503069 ``Quest for unification",
MRTN-CT-2004-005104 ``ForcesUniverse", MRTN-CT-2006-035863
``UniverseNet" and SFB-Transregio 33 ``The Dark Universe" by
Deutsche Forschungsgemeinschaft (DFG). K.C. is supported by the KRF
Grant (KRF-2005-201-C00006) and the KOSEF Grant (KOSEF
R01-2005-000-10404-0) funded by the Korean Government, and also by
the Center for High Energy Physics of Kyungpook National University.
K.C. would like to thank the theory group of Bonn University for the
hospitality during his visit, and B. Acharya and J. P. Conlon for
discussions.

\end{document}